\documentclass{aa}  

\usepackage{graphicx}
\usepackage{txfonts}

\usepackage{hyperref,xcolor,tikz}
\usepackage{colortbl}
\usepackage[switch]{lineno}
\hypersetup{
    colorlinks=true,      
    linkcolor=blue,       
    citecolor=blue,       
    filecolor=magenta,    
    urlcolor=blue,        
    pdfborder={0 0 0}     
}

\newcommand{\nplanets}{1527 }

\newcommand{\orcidicon}{%
    \includegraphics[width=1em]{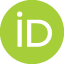}
}
\newcommand{\orcid}[1]{\href{https://orcid.org/#1}{\orcidicon}}
\newcommand{\orcidauthor}[2]{#1\,\orcid{#2}}
\begin{document} 

   \title{Revisiting the conundrum of the sub-Jovian and Neptune desert}
   \subtitle{A new approach that incorporates stellar properties}

   \author{\orcidauthor{C. Magliano }{0000-0001-6343-4744}
          \inst{1,2,3}\thanks{E-mail: \color{blue}christian.magliano@unina.it}         \and
          \orcidauthor{G. Covone}{0000-0002-2553-096X}\inst{1,2,3} \and
          \orcidauthor{E. Corsaro}{0000-0001-8835-2075} \inst{4} \and
          \orcidauthor{L. Inno}{0000-0002-0271-2664}\inst{3,5} \and
          \orcidauthor{L. Cacciapuoti}{0000-0001-8266-0894}\inst{6,7,8}\and
          \orcidauthor{S. Fiscale}{0000-0001-8371-8525} \inst{3,9}\and
          \orcidauthor{I. Pagano}{0000-0001-9573-4928} \inst{8}\and
          V. Saggese\inst{1,2}.}

   \institute{
Dipartimento di Fisica "Ettore Pancini", Università di Napoli Federico II, Napoli, Italy \and
INFN, Sezione di Napoli, Complesso Universitario di Monte S. Angelo,
Via Cintia Edificio 6, 80126 Napoli, Italy  \and
INAF - Osservatorio Astronomico di Capodimonte,
via Moiariello 16, 80131 Napoli, Italy  \and
INAF - Osservatorio Astrofisico di Catania, Via S. Sofia 78, I-95123, Italy \and
Science and Technology Department, Parthenope University of Naples, Naples 80143 Italy  \and
European Southern Observatory, Karl-Schwarzschild-Strasse 2 85748 Garching bei Munchen, Munchen, Germany  \and
Fakultat fúr Physik, Ludwig-Maximilians-Universität München, Scheinerstraße 1, 81679, Munchen, Germany  \and
INAF, Osservatorio Astrofisico di Arcetri, Largo E. Fermi 5, 50125, Firenze, Italy \and
UNESCO Chair ''Environment, Resources and Sustainable Development'', Department of Science and Technology, Parthenope University of Naples, Italy;
}

   \date{Received XXX; accepted YYY}
\titlerunning{A\&A ...}\authorrunning{C. Magliano et al.}        
 
  \abstract
   {The search for exoplanets has led to the identification of intriguing patterns in their distributions, one of which is the  so-called sub-Jovian and Neptune desert. The occurrence rate of Neptunian exoplanets with an orbital period $P\lesssim 4$ days sharply decreases in this region in period-radius and period-mass space.}
   {We present a novel approach to delineating the sub-Jovian and Neptune desert by considering the incident stellar flux $F$ on the planetary surface as a key parameter instead of the traditional orbital period of the planets. Through this change of perspective, we demonstrate that the incident flux still exhibits a paucity of highly irradiated Neptunes, but also captures the proximity to the host star and the intensity of stellar radiation.}
   {Leveraging a dataset of confirmed exoplanets, we performed a systematic analysis to map the boundaries of the sub-Jovian and Neptune desert in the $(F,R_p)$ and $(F,M_p)$ diagrams, with $R_p$ and $M_p$ corresponding to the planetary radius and mass, respectively. 
By using statistical techniques and fitting procedures, we derived analytical expressions for these boundaries that offer valuable insights into the underlying physical mechanisms governing the dearth of Neptunian planets in close proximity to their host stars. }
   {We find that the upper and lower bounds of the desert are well described by a power-law model in the $(F,R_p)$ and $(F,M_p)$ planes. We also obtain the planetary mass-radius relations for each boundary by combining the retrieved analytic expressions in the two planes. 
This work contributes to advancing our knowledge of exoplanet demographics and to refining theoretical models of planetary formation and evolution within the context of the sub-Jovian and Neptune desert.}
   {}


   \maketitle
%
\section{Introduction}
\label{sec:intro}
One of the most intriguing features in the various findings in the realm of exoplanetary science is the scarcity of exoplanets that spans from sub-Jupiters to super-Earths at small orbital distances (period of $\lesssim 4$ days). This is known as the hot sub-Jovian/ Neptune desert (e.g. \citealt{Mazeh2016,Szabo2011}). We adopt this nomenclature here.
\cite{Mazeh2005} first pointed out that a negative linear relation exists between the planetary mass and the orbital period. 
As more planets were discovered in the following years, several other authors examined this tendency (e.g. \citealt{Szabo2011,Beague2013}) and broadened this correlation to the planetary radius (e.g. \citealt{Beague2013, Mazeh2016}). In particular, \cite{Szabo2011} found a sparsely populated region in the $P$-- $M_p$ diagram with $P\leq 2.5$ days, spanning $0.02-0.8\,M_J$, which they termed the sub-Jovian desert. 
Similar findings were also obtained by \cite{Beague2013}, who extended the analysis to the $P$ -- $R_p$ diagram: This diagram shows that planets with $3\,R_\oplus\leq R_p\leq 11R_\oplus$ that orbit their host star in $\lesssim 4$ days are very rare. 
These works pointed out that this region is constrained by two boundaries: an upper boundary delimited by hot Jupiters, and a lower boundary consisting of super-Earths and sub-Neptunes. \cite{Mazeh2016} obtained an analytic description of the two boundaries delimiting the desert in the $\log P$ -- $\log M_p$ and $\log P$ -- $\log R_p$ planes.
Although the debate about the origin of the desert is still ongoing,  the distribution of close-in exoplanets is the result of a non-trivial combination of atmospheric and dynamical processes.
The most accredited scenario today explains the lower boundary by means of the photoevaporation of the H/He atmospheres of an initially low-mass planet \citep{DavisWheatley2009,Lopez2014,OwenLai2018}, while the upper boundary is consistent with numerical simulations based on the high-eccentricity migration scenario and the subsequent tidal angular-momentum exchange of the planet with the star \citep{Benitez2011,Matsakos2016,OwenLai2018,Owen2019}. Photoevaporation is the physical process through which exoplanets shed their H/He envelopes mainly during the first few hundred million years, primarily due to the influence of ultraviolet (UV) and X-ray radiation from their host stars \citep{OwenWu2013}. This reduces the planet size. In this scenario, Neptune-sized planets that either formed within the desert or migrated into it undergo a rapid transformation and become super-Earths. It is worth noting that the lower boundary is influenced by the cumulative UV/X-ray radiation exposure over the planetary lifetime, along with the spectral characteristics of the host star \citep{McDonald2019}. Observational constraints upon the photoevaporative processes that sculpt the desert, or its lower boundary, might come from using the helium as a tracer of escape in UV spectroscopy \citep{Oklopcic2018,Allart2018,Nortmann2018,Salz2018,Allart2019}.  \cite{Guilluy2023} conducted a systematic He I survey of nine exoplanets that belong to the edges of the sub-Jovian and Neptune desert, but found no signature of He absorption at $3\sigma$.
Although the recent findings of \citealt{Thorngren2023} would suggest that photoevaporation still partially causes the formation of the upper boundary, it cannot fully explain the upper edge of the sub-Jovian and Neptune desert as more massive planets ($>0.5\,M_J$) withstand the effects of photoevaporation. 
Numerical simulations by \cite{Ionov2018} and \cite{OwenLai2018} have shown that if photoevaporation were the sole mechanism that sculpts the upper limit, the $P$ -- $R_p$ plane would be populated with sub-Jovian planets that are located at extremely short orbital periods, which contradicts the observed distribution.
Observational evidence that supports these simulations was recently provided by \cite{Vissapragada2022}. In their exploration of helium absorption signals from seven gas-giant planets near the upper boundary, they found that the atmosphere lifetimes of these planets significantly exceed ten billion years. An alternative explanation for the upper limit involves a high-eccentricity migration scenario, in which gravitational interactions can compel a planet into an extremely elliptical orbit that subsequently circularises as the planet draws closer to its parent star. When the orbit is circularised, planets with masses greater than $1\,M_J$ can migrate even closer to their parent star, which is facilitated by the tidal decay caused by the stellar gravitational influence. Observational constraints upon the migration scenario are likely to come from spin-orbit angle measurements of planets lying in proximity of the sub-Jovian and Neptune desert, which is the ultimate aim of the Desert-Rim Exoplanets Atmosphere and Migration (DREAM; \citealt{Bourrier2023}) project. In particular, \cite{Bourrier2023} conducted a homogeneous analysis of $14$ close-in exoplanets using the Rossiter-McLaughlin revolution technique and found $9$ planets on misaligned orbits. In a follow-up paper, \cite{Attia2023} investigated how the spin-orbit misalignments are influenced by the star-planet interactions.

It is generally accepted that the sub-Jovian and Neptune desert cannot be the result of observational biases. Most of the \textit{Kepler} candidates lying in the desert were found to be a false-positive detection \citep{SanchisOjeda2014}. 
However, an increasing number of TESS\footnote{Transiting Exoplanet Survey Satellite.} hot-Neptune planet candidates is spectroscopically confirmed as bona fide planets (e.g. \citealt{Armstrong2020,Burt2020,Murgas2021,Persson2022,Naponiello2023}), as is also shown through the contribution of ground-based transit surveys (e.g. \citealt{Eigmuller2017,West2019,Jenkins2020,Smith2021}). The desert is thus not completely empty, and some authors referred to it as an oasis \citep{Murgas2021} or savanna \citep{Bourrier2023,Szabo2023,Kalman2023}. \cite{CastroGonzalez2024} introduced the term Neptunian ridge to refer to the region of the period-radius space that separates the desert (no planets at the $3\sigma$ level) from the savanna with $3.2\,\text{days}\lesssim P \lesssim 5.7\,\text{days}$.
In addition, \cite{dtarps1} applied a statistical procedure to a sample of $0.9$ million light curves and identified, among other things, hundreds of new TESS candidates that lie within the sub-Jovian and Neptune desert \citep{dtarps2,dtarps3}. If a large fraction of this sample were confirmed by future follow-up efforts, then it would indicate that the desert is not as barren as it appears, but might be caused by an unknown bias in the \textit{Kepler} survey. Nonetheless, the recent vetting efforts of \cite{Magliano2023a,Magliano2023b} based on TESS candidates showed that the false-positive occurrence rate is $\approx 75\,\%$ for candidates in the sub-Jovian and Neptune desert and $\approx 34\,\%$ outside of it. This demonstrates that hot-Neptune candidate planets are twice as likely to be false-positive detections.

In this tangle of conflicting results, a different perspective in which to contextualise this issue might deliver new and interesting cues. \cite{Lundkvist2016} were the first to study the hot-super-Earth desert issue and considered the incident flux on the planetary surface rather than the orbital period. The orbital period is not a completely informative variable for defining a planet as 'hot' because it does not take the characteristics of the host star into account. \cite{Szabo2019} and \cite{Szabo2023} showed that the boundary of the desert depends on the stellar effective temperature, metallicity, and $\log g$, in order of significance. They suggested that the dependence on the fundamental stellar parameters may be an indication of the predominance of photoevaporative processes in determining the boundary, while the non-dependence on tidal forces and $\log g_p$ of the planet is consistent with dynamical processes that partly cause the desert. Thus, the term 'hot' ought to also be correlated to the properties of the star in order to capture the full picture of these systems. Pioneering works that first addressed the sub-Jovian and Neptune desert conundrum (e.g. \citealt{Mazeh2016}) employed the orbital period as the key parameter since it is the most affordable observable. Through the astronomical efforts of different missions (e.g. \textit{Gaia}; \citealt{GaiaCollaboration2016}), a preliminary characterisation of the parent star is almost always feasible today.
For this reason, we study the issue of the sub-Jovian and Neptune desert by replacing the orbital period with the incident flux on the planetary surface. 

The outline of the paper is as follows. In Sect. \ref{sec:methods} we describe our dataset and its features in the $F$ -- $R_p$ and $F$ -- $M_p$ diagrams. In Sect. \ref{sec:results} we then retrieve the analytic expressions of the two boundaries in the two diagrams within a Bayesian framework. Finally, we summarise our conclusions in Sect. \ref{sec:conclusions}.

\section{Methods}
\label{sec:methods}
The insulation flux $F$ corresponds to the amount of stellar energy received per unit area on the planetary surface and is defined as
\begin{equation}
\nonumber
F=\dfrac{L_*}{4\pi d^2},
\end{equation}
where $L_*$ is the luminosity of the star, and $d$ is the distance of the planet from the star (semi-major axis of the orbit). For context, Earth's insulation flux $F_\oplus$ is approximately $1361\,\text{W/}\text{m}^2$, and it provides a useful reference point for comparing the energy received by other planets.
However, the insulation flux $F$ that a planet receives from its host star (modelled as a black-body source) in units of the Earth value $F_\oplus$ in the circular orbit approximation can also be expressed as
\begin{equation}
    \dfrac{F}{F_\oplus}=\left(\dfrac{\rho_*}{\rho_\odot}\right)^{-2/3}\left(\dfrac{P}{1\text{yr}}\right)^{-4/3}\left(\dfrac{T_*}{T_\odot}\right)^{4} \, ,
    \label{eq:flux_definition}
\end{equation}
where $\rho_*$ is the stellar mean density, $\rho_\odot=1.408\, {\rm g/cm}^3$ is the mean density of the Sun, $P$ is the orbital period of the planet, $T_*$ is the stellar effective temperature, and $T_\odot=5778\,\text{K}$ is the effective temperature of the Sun. Thus, the insulation flux simultaneously takes the orbital period of the planet and the host star temperature and mean density into account. At a fixed orbital period, the radiation received by the planet grows when it orbits a hotter and less dense star. 

We initially used a dataset that consisted of 3335 confirmed exoplanets\footnote{https://exoplanetarchive.ipac.caltech.edu/index.html} with a known orbital period and planetary radius that orbit a star with an estimated mass and radius. The corresponding stellar parameters were retrieved from the NASA Exoplanet Archive. The requirement of a known planetary radius implies that approximately $99\%$ of the planets in the sample were discovered using the transit method. This observational bias likely results in an overrepresentation of larger, short-period exoplanets, particularly those orbiting bright stars. While this bias does not significantly affect the primary findings of this work, such as the observed paucity of highly irradiated Neptune- and sub-Jovian-sized planets, it may influence the accuracy of the boundary calculations at lower flux values, where smaller or more distant planets are potentially underrepresented. Future exoplanetary missions that aim to expand the demographic reach of exoplanet detection will be crucial in strengthening and refining the accuracy of these results.

We removed $1808$ ($\approx 54\,\%$ of the total) exoplanets from this initial sample for one of the following reasons: \textit{i}) Their radius or period estimates had an uncertainty larger than $10\%$, or \textit{ii}) they orbit stars whose radius or mass estimates have an uncertainty larger than $10\%$. After this filtering, our final sample ($S_1$ hereafter) was composed by \nplanets exoplanets. We used a threshold value of $10\%$ for the maximum uncertainty in order to minimise the propagated error over the insulation flux and to have the cleanest and most complete dataset possible. Only $0.6\%$ of the planets in $S_1$ have $\Delta F/F>40\%$, and none exceeds $50\%$. 

\subsection{Orbital period against incident flux}
\label{sec:dicotomy}

\begin{figure*}
    \centering
    \includegraphics[width=0.95\textwidth]{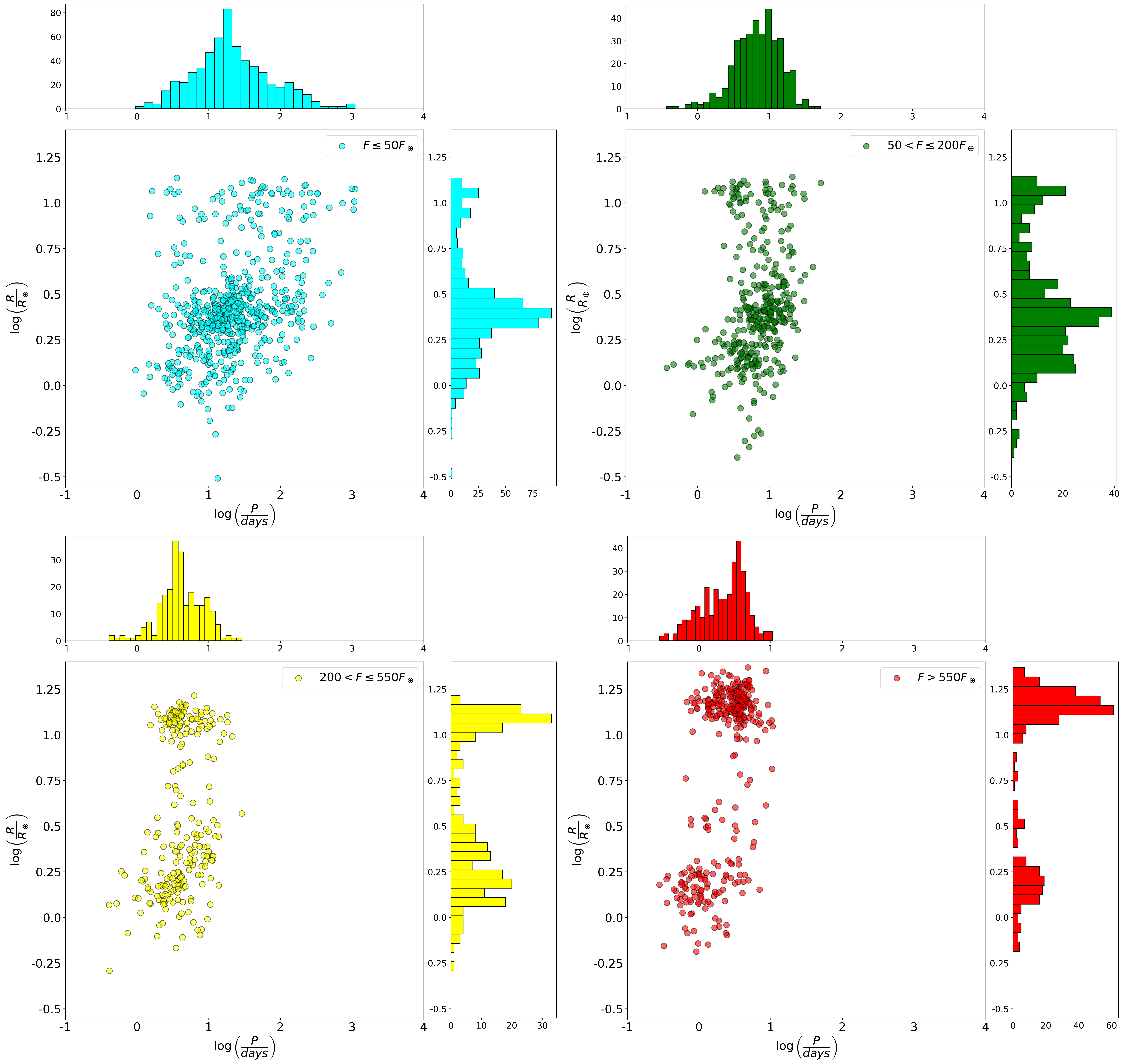}
    \caption{Distribution of the confirmed exoplanets of the $S_1$ sample in the $\log P$ -- $\log R_p$ diagram for different ranges of the insulation flux. The \textit{ upper left} panel shows the $586$ planets that receive $\leq 50\,F_\oplus$, the \textit{upper right} panel shows the $364$ planets that receive between $50\,F_\oplus$ and $200\,F_\oplus$, the \textit{lower left} panel depicts the $238$ planets that receive between $200\,F_\oplus$ and $550\,F_\oplus$, and the \textit{lower right} panel illustrates the $339$ planets with $F>550\,F_\oplus$.}
    \label{fig:beyond_orbital_period}
\end{figure*}

In order to investigate whether $F$ is a more appropriate physical quantity to describe the sub-Jovian and Neptune desert, we preliminary divided the $S_1$ sample into four groups of exoplanets: \textit{i}) the first group was composed of $586$ planets that received an insulation flux $F<50\,F_\oplus$, \textit{ii}) the second group consisted of  $364$ planets whose incident flux was between $50\,F_\oplus$ and $200\,F_\oplus$, \textit{iii}) $238$ planets in the third group had $200\,F_\oplus<F\leq 550\,F_\oplus$, and \textit{iv}) the fourth group comprised the remaining $339$ planets with $F>550\,F_\oplus$. We emphasise that the flux binning we used in this step was completely arbitrary and was simply provided as a guideline to distinguish between extremely and moderately irradiated objects. The selection of four bins 
aimed to balance the need for granularity in the analysis with the necessity of maintaining sufficient statistical power within each bin. We ensured that each bin contained a sufficient number of data points that allowed us to apply robust statistical methods.
In Fig. \ref{fig:beyond_orbital_period} we depict the distribution of the four groups in the $P$ -- $R_p$ plane. This chart would suggest that the less irradiated exoplanets are more uniformly distributed than the highly irradiated ones. 
At $F<50\,F_\oplus$, there is no evidence of a sub-Jovian and Neptune desert. However, as $F$ increases, a gap opens in the distribution, and the Neptune/sub-Jupiter region in $R_p$ is voided of planets. 
The lower left and right panels ($200\,F_\oplus<F\leq 550\,F_\oplus$ and $F>550\,F_\oplus$, respectively) show a clear dichotomy of highly irradiated exoplanets: hot Jupiters or hot super-Earths/sub-Neptunes. It therefore appears that the insulation flux plays the main role in sculpting the sub-Jovian and Neptune desert, while the orbital period is an incomplete physical parameter, as we show in Sect. \ref{sec:results}. We also point out the well-known lack of inflated Jupiters ($R_p\gtrsim 15\,R_J$) that receive less than $550\,F_\oplus$ \citep{Laughlin2011,Miller2011,Demory2011,Laughlin2015,Thorngren2018}. These planets, similar to Jupiter in composition, appear to be puffed up and less dense. The cause of this inflation is not fully understood, but may involve factors such as stellar irradiation, internal heat, and tidal forces.
In order to statistically demonstrate the differences among the four aforementioned planetary populations, we employed a two-sample Kolmogorov-Smirnov test (e.g. \citealt{Feigelson2012}) on the planetary radius distributions of these groups. Specifically, for each pairing of the four datasets, we derived a p value lower than $10^{-15}$, which conclusively establishes the statistical disparity in the planetary radius distributions in all four groups.

We concluded this preliminary analysis of our dataset by replicating the steps performed by \cite{Szabo2019} on our $S_1$ sample, which is nearly three times larger than the sample they used, despite the stricter constraints we adopted on the uncertainties over each parameter. \cite{Szabo2019} employed the two-sample Kolmogorov-Smirnov test to assess differences in the planetary distributions in the $P$ -- $R_p$ diagram according to a third variable (e.g. stellar mass, radius, or metallicity). By comparing the characteristics of exoplanets inside and outside the desert region, the study shed light on the factors that contribute to its formation.
We confirmed the dependence of the desert edges in the $P$ -- $R_p$ diagram on the stellar effective temperature, mass, and metallicity found by \cite{Szabo2019}. In particular, we recovered that planets in the desert are likely to have a shorter period around cooler, less massive stars and more metal-rich stars, as obtained by \cite{Szabo2019}. However, the role of the metallicity in sculpting the desert was statistically downsized by the latest work of \cite{Szabo2023} based on a new dataset in which the stellar parameters were uniformly retrieved in the Apache Point Observatory Galactic Evolution Experiment (APOGEE) DR17 sample \citep{Majewski2017,Zasowski2017}.

\subsection{The sub-Jovian/Neptunian desert in the flux-radius plane}
\label{sec:flux_rad_analysis}
\begin{figure*}
    \centering
    \includegraphics[width=0.95\textwidth]{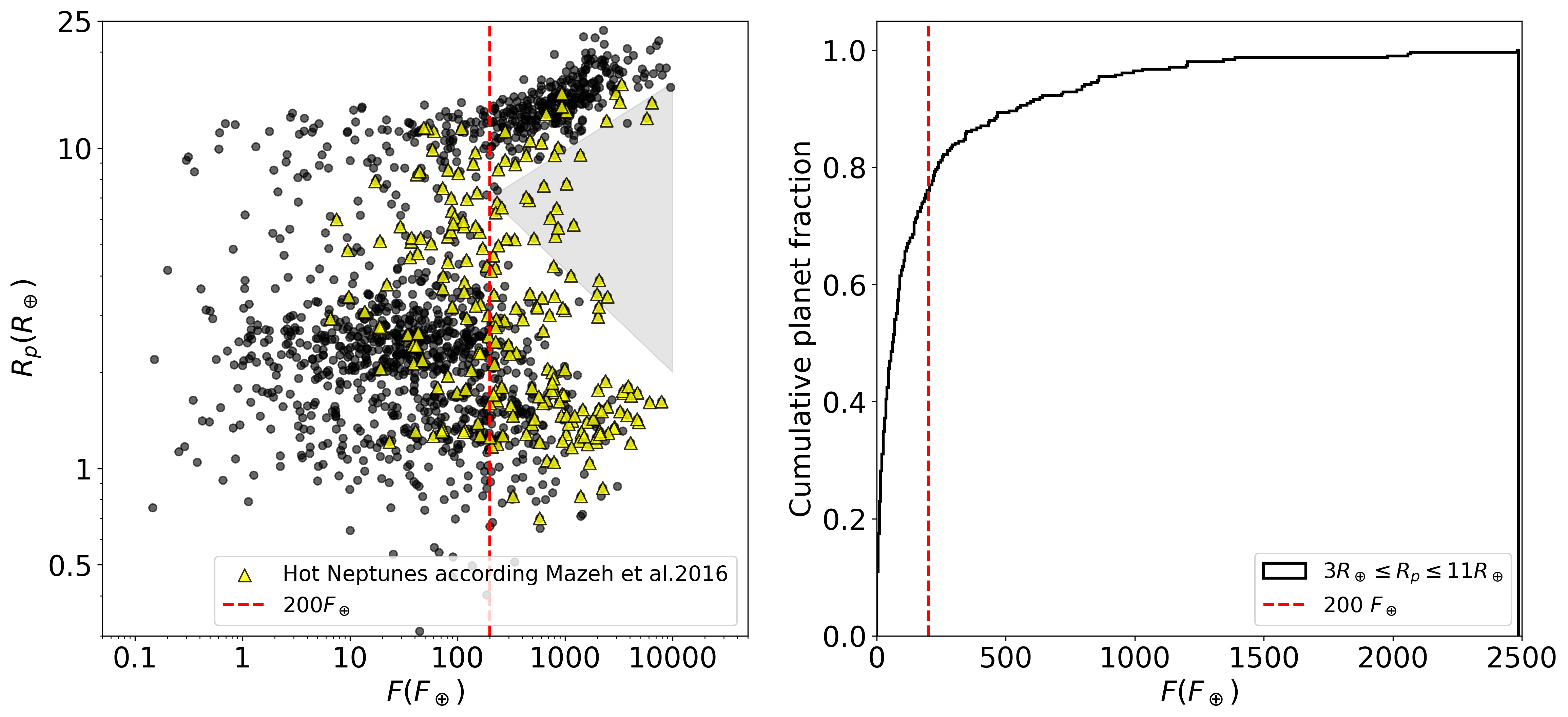}
    \caption{Analysis of the planetary distribution in the $F$ -- $R_p$ plane and the flux cumulative distribution function. \textit{Left panel}: Distribution of the \nplanets planets of the $S_1$ set in the $F$ -- $R_p$ plane. The yellow triangles represent planets that would be referred to as hot Neptunes according to the limits obtained by \citealt{Mazeh2016}. The grey triangle is a qualitative representation of the desert. \textit{Right panel}: Cumulative distribution function of incident flux values $F$ of planets with $3R_\oplus\leq R_p\leq 11 R_\oplus$. The vertical dashed line in both panels represents the threshold value of $200\,F_\oplus$ below which lie $75\%$ of the planets with $3R_\oplus\leq R_p\leq 11 R_\oplus$.}
    \label{fig:desert_flux_rad}
\end{figure*}

In the left panel of Fig. \ref{fig:desert_flux_rad}, we display the distribution of the \nplanets planets in $S_1$ in the $F$ -- $R_p$ plane. This parameter space shows a scarcity of Neptune- and sub-Jupiter-like planets that receive at least $\approx 200\, F_\oplus$ and form a vaguely triangular region that is delimited by two boundaries, as delineated in grey in Fig. \ref{fig:desert_flux_rad} based on visual inspection.
The lower edge comprises highly irradiated super-Earths and sub-Neptunes, and the upper edge consists of hot Jupiters. 
Furthermore, our sample contains $221$ planets that lie between the boundaries of the desert obtained by \citealt{Mazeh2016} in the $P$ -- $R_p$ plane. These objects would be referred to as hot Neptunes if we only considered their distribution in the $P$ -- $R_p$ plane, but most of them are found in crowded regions in the $F$ -- $R_p$ plane, thus leaving the desert (see Sect. \ref{sec:diamonds_rad}).

\begin{figure}
    \centering
    \includegraphics[width=0.47\textwidth]{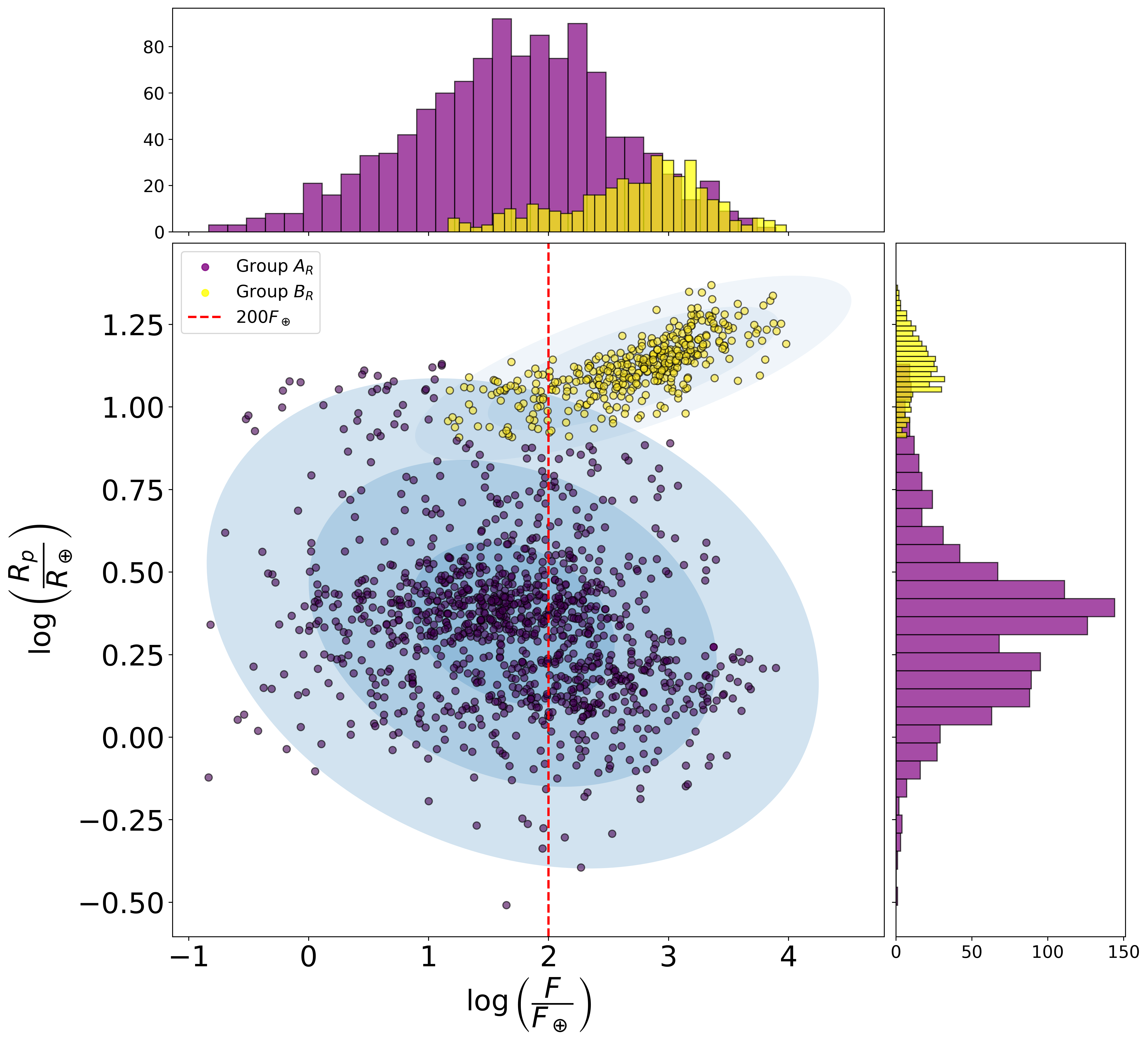}
    \caption{Application of a two-component bivariate GMM over $S_1$ sample in the $\log F$ -- $\log R_p$ space. We found two separate populations: Group $A_R$ (purple dots) with a mean value of $(1.70, 0.34)$, and Group $B_R$ (yellow dots) with a mean value of $(2.72, 1.12)$. }
    \label{fig:GMM_2_component}
\end{figure}

As shown in the right panel of Fig. \ref{fig:desert_flux_rad}, planets with $3 \, R_\oplus\leq R_p\leq 11 \, R_\oplus$ are not so rare when the flux received from their host star is $\lesssim 200\,F_\oplus$. In particular, Neptunes and sub-Jupiters that experience an incident flux $F\leq 200F_\oplus$ are three times more common than those with $F > 200 \, F_\oplus$. 
In order to describe the sub-Jovian and Neptune desert in this new parameter space, we studied the radius distribution of the planets in $S_1$ in terms of the received insulation flux.
The left panel of Fig. \ref{fig:desert_flux_rad} suggests that there are two distinct populations of exoplanets: \textit{i}) sub-Neptunes and super-Earths with an incident flux spanning almost five orders of  magnitudes, and \textit{ii}) Jupiters and inflated hot Jupiters that are found to withstand more stellar radiation on average. 
We applied a Gaussian mixture model (GMM; \citealt{Ivezic2020}) to distinguish between the two classes of planets. A GMM is a probabilistic model that is used for clustering and a density estimation. It assumes that the data points are generated from a mixture of several Gaussian distributions, each characterised by its mean and covariance. For the multivariate case, a GMM with $K$ components is described by a density probability $p(\Vec{x})$ given by 
\begin{equation}
    p(\Vec{x})=\sum_{i=1}^{K}\phi_i\mathcal{N}(\Vec{x}\vert \Vec{\mu}_i,\Sigma_i),
\end{equation}
where $\Vec{\mu}_i$ and $\Sigma_i$ correspond to the mean and the covariance of the $i$th component, respectively, $\mathcal{N}$ refers to a Gaussian density probability, and $\phi_i$ is the mixture component weight of the $i$th component, such that $\sum_{i=1}^{K}\phi_i=1$. We calculated the best parameters for each component by using the expectation-maximisation algorithm \citep{Meng2002}. The expectation stage calculates the probability of each data point belonging to each Gaussian component, and the maximisation step fine-tunes the model parameters based on these probabilities.
We performed a two-component (i.e. $K=2$) bivariate GMM in the $\log F$ -- $\log R_p$ space in order to cluster $S_1$ into two groups, as shown in Fig. \ref{fig:GMM_2_component}. 
We opted to work with the log values since the planetary fluxes are spread over a wide range of magnitudes. 
Figure \ref{fig:GMM_2_component} clearly shows two well-separated class of exoplanets. The first component of the model, group $A_R$ hereafter, consists of a variety of exoplanet types: super-Earths/sub-Neptunes, hot Neptunes/Saturns, and cold giants. We stress that there is still a less dense (but not empty) populated region of highly irradiated sub-Jovians and Neptunes that  would represent the savanna, similar to what we found in the $P$ -- $R_p$ and $P$ -- $M_p$ diagrams.
We found the mean of group $A_R$ to be located at $\Vec{\mu}_{A_R}\simeq(50\,F_\oplus, \, 2\,R_\oplus)$. 
On the other hand, the second component corresponds to hot Jupiters and inflated hot Jupiters, group $B_R$ hereafter. The mean of group $B_R$ corresponds to $\Vec{\mu}_{B_R}\simeq(523\,F_\oplus,13\,R_\oplus)$. The GMM allowed us to label each point of our diagram as belonging to one of the two groups. Since the focus of this work is on highly irradiated exoplanets, we only considered the bodies that received at least $200\,F_\oplus$.  Thus, we took the $75$th percentile of the incident flux distribution of planets with $3\,R_\oplus\leq R_p\leq 11\,R_\oplus$ as the threshold value above which the sub-Jovian and Neptune desert is defined.

\begin{figure}
    \centering
\includegraphics[width=0.48\textwidth]{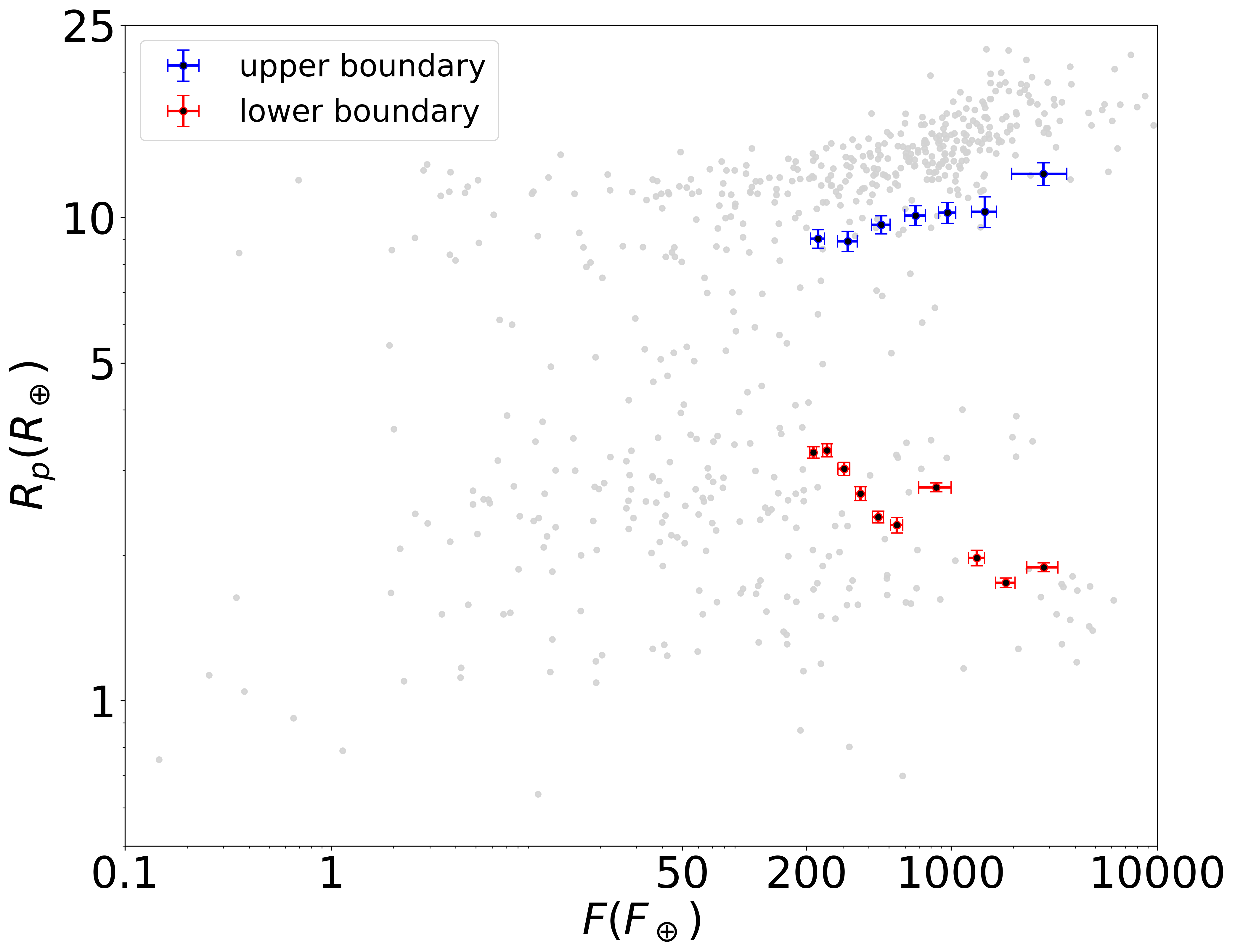}
    \caption{Calculated boundaries in $(F, R_p)$ space: $(F_{\text{low},i},R_{\text{low},i})$ pairs (purple points) sample the lower boundary, and $(F_{\text{up},i},R_{\text{up},j})$ pairs (red points) mark the upper boundary.}
    \label{fig:bounds_desert}
\end{figure}

The next step was to determine the borders in the $\log F$ -- $\log R_p$ diagram that define the lower and upper edges of the desert. In particular, we defined the edges of the two groups with a quantitative criterion, and the results are shown in Fig. \ref{fig:bounds_desert}. 
We sliced the diagram in terms of the incident flux in order to study the planetary radius distribution of groups $A_R$ and $B_R$ in each slice. For both groups, we defined a custom grid of $F$ values using a binning that ensured at least $20$ elements per group. This approach allowed us to calculate statistically reliable uncertainties for the representative value of each group. In particular, the main purpose of the grid-based analysis was to calculate the points that were used below to derive the borders (see Sect. \ref{sec:diamonds_rad}).
Group $A_R$ is slightly larger than group $B_R$, and its grid had ten slices, while the grid of group $B_R$ had seven slices. The two grids span $\left[200\,F_\oplus, 4000\,F_\oplus\right]$, which provides enough points for a statistical analysis. Moreover, as shown in Fig. \ref{fig:GMM_2_component}, the $R_p$ distribution of group $A_R$ is skewed towards larger radii by the cold and warm sub-Jovians/Jupiters. We also applied a Shapiro-Wilk test \citep{Shapiro1965} to verify that it is not normally distributed. We rejected the null hypothesis with $p<10^{-13}$. In particular, the scarcity of warm and hot sub-Jovians belonging to group $A_R$, within the savanna and nearby the exterior ellipsoids of group $B_R$, would still contaminate the lower border calculation. 
In order to avoid contamination from hot sub-Jovians of group $A_R$ inside the desert, we removed them by performing a $2\sigma$ clipping for the $R_p$ distribution of group $A_R$ in each slice.

We defined the lower boundary point in the $i$th slice of the flux grid related to group $A_R$ as
\begin{equation}\label{eq:r_low}
    R_{\text{low},i}=\mu_{A_R,i}+2\sigma_{A_R,i}\qquad i=1,..,10 \, ,
\end{equation}
where $\mu_{A_R,i}$ and $\sigma_{A_R,i}$ represent the mean and the standard deviation of the the $R_p$ distributions of group $A_R$ in the $i$th slice, respectively. 
The incident flux $F_{\text{low},i}$, associated with $R_{\text{low},i}$, was taken as the average flux within the $i$th slice. The uncertainties $\Delta R_{\text{low},i}$ and $\Delta F_{\text{low},i}$ on the mean planet radius and incident flux, respectively, were estimated as the error on the standard deviation of the $R_p$ distribution for group $A_R$  and the standard deviation of the $F$ distribution for group $A_R$, both evaluated for the $i$th slice. We note that if we had not employed the sigma-clipping process, $\sigma_{A_R,i}$ in each slice would have been affected by the presence of a few hot sub-Jovians and would thus have yielded a point $(F_{\text{low},i},R_{\text{low},i})$ that  would have fallen inside the desert instead of at its lower edge. This methodological difficulty is caused by the Neptunian ridge \citep{CastroGonzalez2024} in the flux-radius diagram that separates the desert from the underpopulated savanna.

We applied an analogous treatment for the planets falling within group $B_R$ in order to build up the upper boundary of the desert. We did not employ any sigma clipping since no planet of group $B_R$ contaminated the upper edge calculation. In this case, the upper boundary in planet radius is defined as
\begin{equation}\label{eq:r_up}
R_{\text{up},j}=\mu_{B_R,j}-2\sigma_{B_R,j}\qquad j=1,... 7 \, .
\end{equation}

The decision to define the boundary points as the points with a $z-$score equal to $2$ in each distribution is arbitrary, but was made in light of an a posteriori visual analysis. As shown by the confidence ellipsoids in Fig. \ref{fig:GMM_2_component}, if we used $z=3$ or $z=1$, the points would be significantly farther out from the edges on the outside and inside, respectively. In particular, in each slice, more than $90\%$ of exoplanets of groups $A_R$ and $B_R$ are found to be above (below) the upper (lower) boundary.
Moreover, although at least $20$ objects in each bin might be an arbitrary choice, this was the best trade-off between having solid statistical results and a sufficient number of boundary points to later fit with high-DImensional And multi-MOdal NesteD Sampling (\textsc{Diamonds}, \citealt{Corsaro2014}). When we decreased the number of objects per bin to 10 in order to almost duplicate the number of boundary points, those describing the lower edge were mostly unreliable. This is mainly due to contaminants in the savanna whose weights in the boundary point computation are not attenuated by the poor statistics.

We observed that the $R_p$ distribution of the super-Earths and sub-Neptunes (group $A_R$) slowly narrows and shifts towards a smaller mean planetary radius as $F$ increases. On the other hand, the $R_p$distribution of the giant planets (group $B_R$) broadens and quickly tends to a larger mean planetary radius when $F$ increases. 
These two effects at high fluxes lead to the observed dichotomy of highly irradiated exoplanets and to the peculiar triangular shape of the sub-Jovian and Neptune desert.
In Fig. \ref{fig:bounds_desert} we show the points that delimit the lower and upper edges in $(F, R_p)$ space within the range $[200\,F_\oplus,4000\,F_\oplus]$.

\subsection{The sub-Jovian/Neptunian desert in the flux-mass plane}
\label{sec:flux_mass_analysis}

\begin{figure*}
    \centering
    \includegraphics[width=0.95\textwidth]{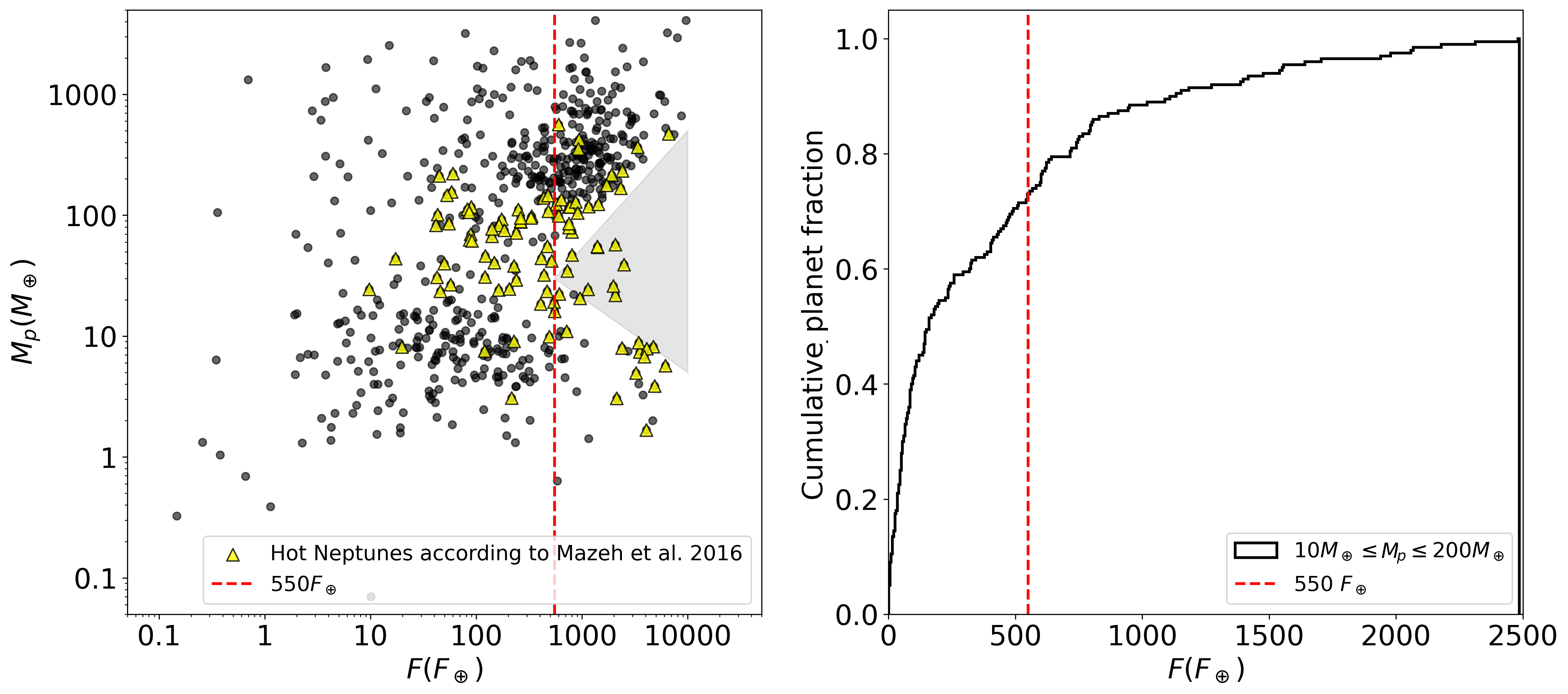}
    \caption{Analysis of the planetary distribution in the $F$ -- $M_p$ plane and the flux cumulative distribution function. \textit{Left panel}: Distribution of the 609 planets of the $S_2$ set in the $F$ -- $M_p$ plane. The yellow triangles represent planets that would be referred to as hot Neptunes according to the limits obtained by \cite{Mazeh2016}. The grey triangle is a qualitative representation of the sub-Jovian and Neptune desert. \textit{Right panel}: Cumulative distribution function of incident flux values $F$ of planets with $10 M_\oplus\leq M_p\leq 200 M_\oplus$. The vertical dashed line represents the threshold value of $550 \,F_\oplus$ in both panels below which $75\%$ of the planets with $10 M_\oplus\leq M_p\leq 200 M_\oplus$ lie.}
    \label{fig:desert_flux_mass}
\end{figure*}

Within our initial sample, 609 planets out of \nplanets also have a planetary mass estimate with $\Delta M_p/M_p<30\,\%$. We refer to this set as sample $S_2$. We stress that the higher threshold value for $\Delta M_p/M_p$ with respect to $\Delta R_p/R_p$ was set to guarantee enough statistics in the grid-analysis step. If we had required a $\Delta M_p/M_p<10\,\%$, it would have halved the $S_2$ sample.
In Fig. \ref{fig:desert_flux_mass} we show the distribution of the objects belonging to the $S_2$ sample in the $F$ -- $M_p$ plane. 
In this parameter space, we still observe a paucity of Neptunes and sub-Jupiters that receive at least $\approx 550\, F_\oplus$. In this case as well, only a few Hot Neptunes, defined as the objects within the boundaries of  \citealt{Mazeh2016}, lie within the underpopulated region (see Sect. \ref{sec:diamonds_mass}).
The right panel of Fig. \ref{fig:desert_flux_mass} shows that planets with $10 \, M_\oplus\leq M_p\leq 200 \, M_\oplus$ are common when the flux received from their host star is $\lesssim 550\,F_\oplus$. Neptunes and sub-Jupiters that experience an incident flux $F\leq 550F_\oplus$ are three times more frequent than those with $F > 550 \, F_\oplus$. 
Therefore, in contrast to the flux-radius diagram, the vertex of the triangle (namely the $75$th percentile of the incident flux distribution of planets with mass $10 \, M_\oplus\leq M_p\leq 200 \, M_\oplus$) here appears to be shifted towards higher incident flux (see Sect. \ref{sec:flux_threshold}). As done in Sec. \ref{sec:dicotomy}, we confirmed all the results found by \cite{Szabo2019} and \cite{Szabo2023} in the $P$ -- $M_p$ plane, with planets in the desert likely to have a shorter period around cooler and more metal-rich stars. We also recovered their finding that the stellar mass does not play a critical role in sculpting the desert in the $P$ -- $M_p$ diagram.

\begin{figure}
    \centering
    \includegraphics[width=0.47\textwidth]{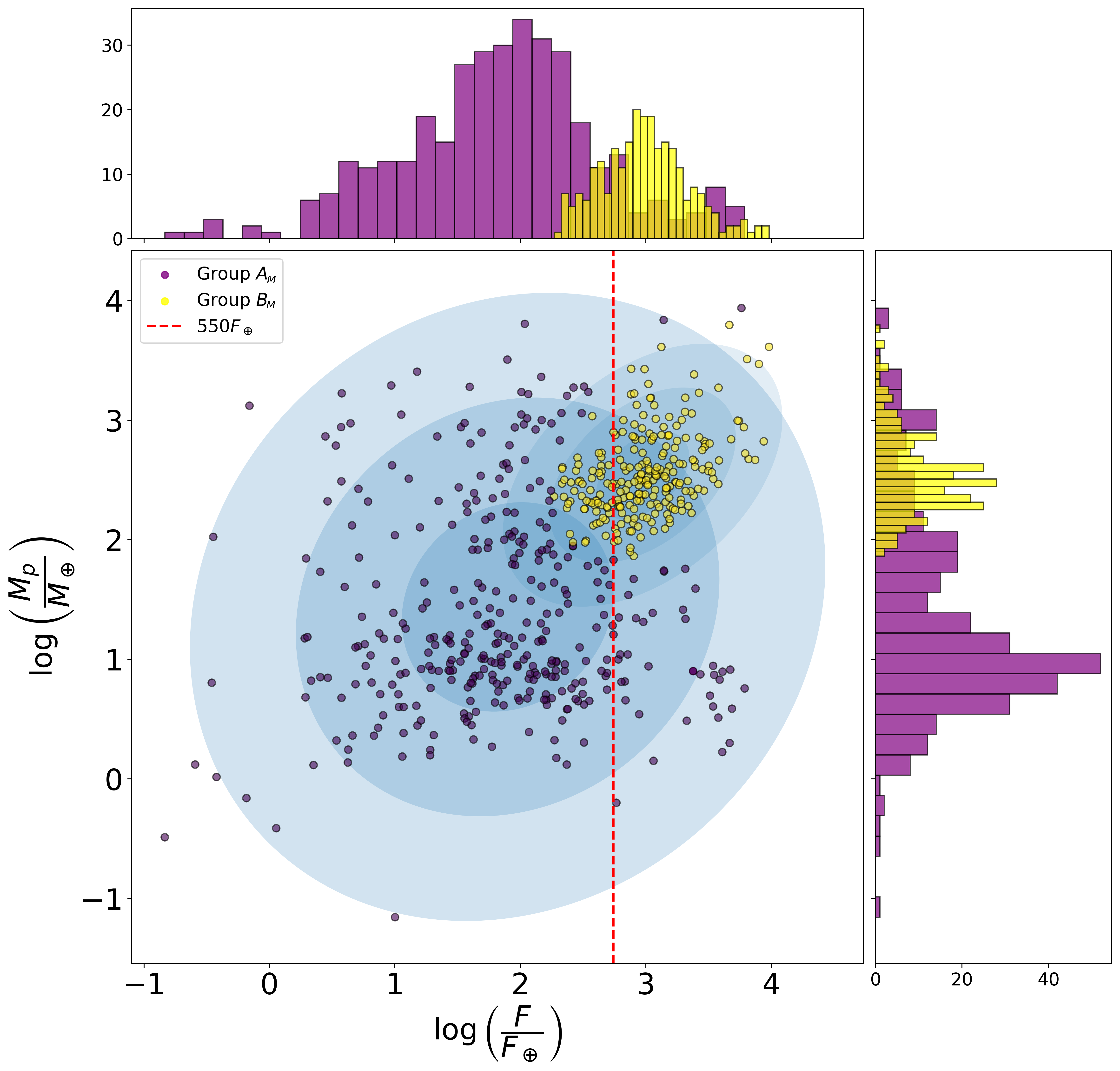}
    \caption{Application of a two-component bivariate GMM on the sample $S_2$ in the $\log F$ -- $\log M_p$ space. We identified two separate populations: Group $A_M$ (purple dots) with a mean value of $(1.8, 1.4)$, and group $B_M$ (yellow dots) with a mean value of $(2.9, 2.5)$. }
    \label{fig:GMM_2_component_mass}
\end{figure}

We repeated the same steps as in Sect. \ref{sec:flux_rad_analysis} considering the planetary mass in order to find the analytic expression of the two boundaries in the $(F, M_p)$ space. 
We report the result of the two-component bivariate GMM in $\log F$ -- $\log M_p$ space in Fig. \ref{fig:GMM_2_component_mass}. Group $A_M$ consists of a wide variety of exoplanet types spanning several orders of magnitude in incident flux (from $\approx 0.1\,F_\oplus$ to $\approx 10^4\,F_\oplus$) as well as a broad spectrum of masses (from $\approx 0.1\,M_\oplus$ to $\approx 8000\,M_\oplus$). We also point out that the most massive objects in group $A_M$ tend to experience low levels of incident flux. We found the mean of group $A_M$ to be located at $\Vec{\mu}_{A_M}\simeq(78\,F_\oplus,27\,M_\oplus)$. On the other hand, group $B_M$ corresponds to planets that are more massive and more irradiated than those in group $A_M$ on average. We found the mean of group $B_M$ to be located at $\Vec{\mu}_{B_M}\simeq(939\,F_\oplus,347\,M_\oplus)$. It also covers a narrower flux and mass ranges, as shown by the confidence bands of the ellipsoids. However, group $B_M$ still spans two orders of magnitudes in planetary mass beyond the threshold flux of $550\,F_\oplus$, and it is not Gaussian (Shapiro-Wilk test, $p<10^{-6}$) but a skewed distribution towards more massive planets, as shown by the marginal histogram in Fig. \ref{fig:GMM_2_component_mass}.

Thus, we needed to remove extremely massive exoplanets that otherwise would lead to large spreads in the giant populations and would distort the upper boundary computation. 
For this reason, we discarded all the planets with a planetary mass $M_p>600\,M_\oplus$,  thus resulting in a final sample composed of $512$ exoplanets with a planetary mass estimate ($S_3$ sample hereafter). The choice of using a cutoff value of $600\,M_\oplus$ was arbitrary, but is capable of guaranteeing enough statistics in each slice.

This operation does not affect the robustness of our results since we are interested in describing the low-mass edge of the group $B_M$ distribution. However, the $S_3$ sample contains fewer than half of the objects that were contained in the initial sample we used in Sect. \ref{sec:flux_rad_analysis}, so that the statistical analysis yields larger errors over the upper and lower boundary points. 
Similarly, the fitted parameters through \textsc{Diamonds} are more uncertain than those retrieved in Sect. \ref{sec:flux_rad_analysis}. 
A summary of the three samples defined in this work is given in Table \ref{tab:tab_samples}

\begin{table*}
\centering
\caption{Samples defined in this work and their description.}
\begin{tabular}{cccc}
\hline
\multicolumn{1}{c|}{Name}  &    \# planets & Description                                                                                             & Notes     \\ \hline
\multicolumn{1}{c|}{$S_1$} & 1527       & Planets with $\Delta R_p/R_p, \Delta P/P, \Delta M_*/M_*,\Delta R_*/R_*$ and $\Delta T_*/T_*<10\%$ & Sect. \ref{sec:methods} \\
\multicolumn{1}{c|}{$S_2 \left(\subset S_1\right)$} &609     & $S_1$ planets with known mass $\left(\Delta M_p/M_p<30\%\right)$                                                                & Sect. \ref{sec:flux_mass_analysis} \\
\multicolumn{1}{c|}{$S_3 \left(\subset S_2\right)$}& 512 &  $S_2$ planets with $M_p<600\,M_\oplus$                                                                           & Sect. \ref{sec:flux_mass_analysis} \\ \hline
\end{tabular}
\label{tab:tab_samples}
\tablefoot{The column notes indicate where the corresponding sample is defined.}
\end{table*}

For the boundary calculations, we only considered exoplanets that received at lest $550\, F_\oplus$. For both groups in our sample, we created a custom grid of $F$ values spaced such that each slice contained at least $20$ elements per group. As anticipated, since the sample is much less numerous in this space of parameter, we needed to choose larger slices than we defined in Sect. \ref{sec:flux_rad_analysis}. 
For this reason, the grid used to sample group $A_M$ consisted of only three slices and that used to slice group $B_M$ contained just four slices. The two grids span the range $\left[550\,F_\oplus,4000\,F_\oplus\right]$, and the upper limit was set to ensure enough points for a statistical analysis. In this case as well, the contamination of highly irradiated sub-Jovian planets within the savanna forced us to perform a $2\sigma$ clipping for the $M$ distribution of group $A_M$ in each slice. 

In contrast with the boundary definitions given in Eqs. \eqref{eq:r_low}-\eqref{eq:r_up}, since the planetary masses were spread over several orders of magnitudes, we defined the boundary points as points with a $z$-score equal to one, as suggested by the confidence ellipsoids in Fig.\ref{fig:GMM_2_component_mass}. In this case as well, the choice of the proper $z$ score was based on an a posteriori visual inspection of the points delimiting the two boundaries. However, due to the poorer statistics in each slice, more than $75\%$ of the exoplanets in groups $A_M$ and $B_M$ were found to be above (or below) the upper (or lower) boundary.

\begin{figure}
    \centering
\includegraphics[width=0.48\textwidth]{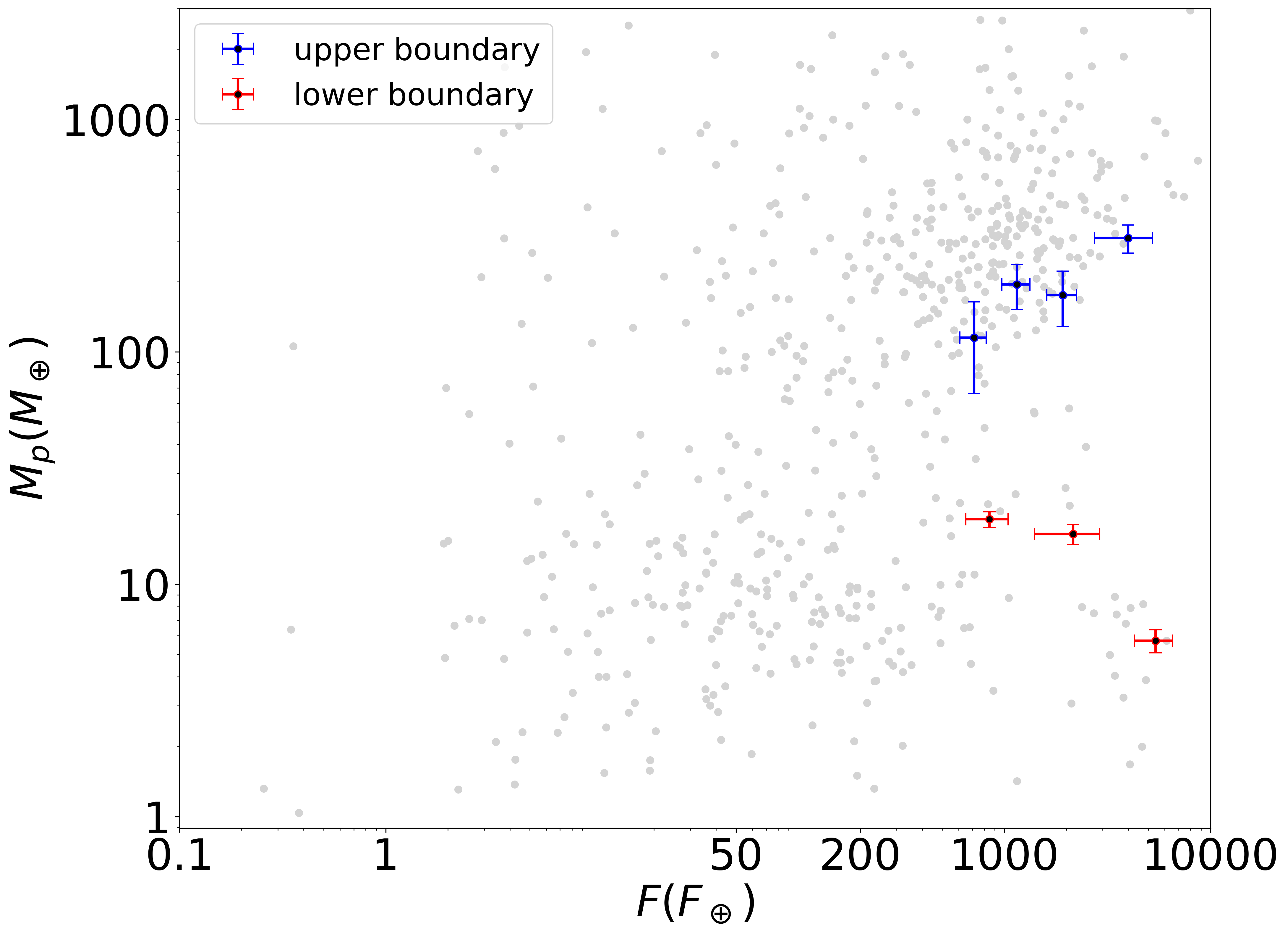}
    \caption{Calculated boundaries in the $(F, M_p)$ space: $(F_{\text{low},i},M_{\text{low},i})$ pairs (purple points) sample the lower boundary, and $(F_{\text{up},i}, M_{\text{up},j})$ pairs (red points) mark the upper boundary of the desert.}
    \label{fig:bounds_desert_mass}
\end{figure}

In particular, the lower boundary point in the $i$th slice of the flux grid related to group $A_M$ was calculated as
\begin{equation}\label{eq:r_low_mass}
    M_{\text{low},i}=\mu_{A_M,i}+\sigma_{A_M,i}\qquad i=1,..,3 \, .
\end{equation}
 The upper boundary point in the $j$th slice of the flux grid related to group $B_M$ is given by
\begin{equation}\label{eq:r_up_mass}
    M_{\text{up},j}=\mu_{B_M,j}-\sigma_{B_M,j}\qquad j=1,...,4 \, .
\end{equation}
In Fig. \ref{fig:bounds_desert_mass} we show the points that delimit the lower and upper boundary in the $(F,M_p)$ space within the range $[550F_\oplus,4000F_\oplus]$. 

\subsection{About the threshold incident flux}\label{sec:flux_threshold}
In the $F$ -- $R_p$ and $F$ -- $M_p$ planes (see Sects. \ref{sec:flux_rad_analysis} and \ref{sec:flux_mass_analysis}), we defined the threshold incident flux from which to start our analysis as the $75$th percentile  of the distribution of exoplanets with $3\,R_\oplus\leq R_p\leq 11\,R_\oplus$ and $10\,M_\oplus\leq M_p \leq 200\, M_\oplus$, respectively. We refer to this value as $F_{75\text{th}}$. 
We found  $F_{75\text{th}}\approx 200\,F_\oplus$ in the $F$ -- $R_p$ plane and  $F_{75\text{th}} \approx 550\,F_\oplus$ in the $F$ -- $M_p$ plane. 
This raises the intriguing question whether the variation in $F_{75\text{th}}$ between $F$ -- $R_p$ and $F$ -- $M_p$ reflects an underlying physical cause, or if is it merely a consequence of the sparser data in the mass plane.
In order to address the second option, we employed two different strategies: \textit{i)} We reduced the size of the $S_1$ sample to match that of $S_2$ and calculated $F_{75\text{th}}$ for planets with $3\,R_\oplus \leq R_p \leq 11\,R_\oplus$, and  \textit{ii)} we enlarged the size of the $S_2$ sample to match that of $S_1$ and calculated $F_{75\text{th}}$ for planets with $10\,M_\oplus \leq M_p \leq 200\,M_\oplus$.

\subsubsection{Downsizing the $S_1$ sample}\label{sec:decrease_S1}

\begin{figure}
    \centering
    \includegraphics[width=\linewidth]{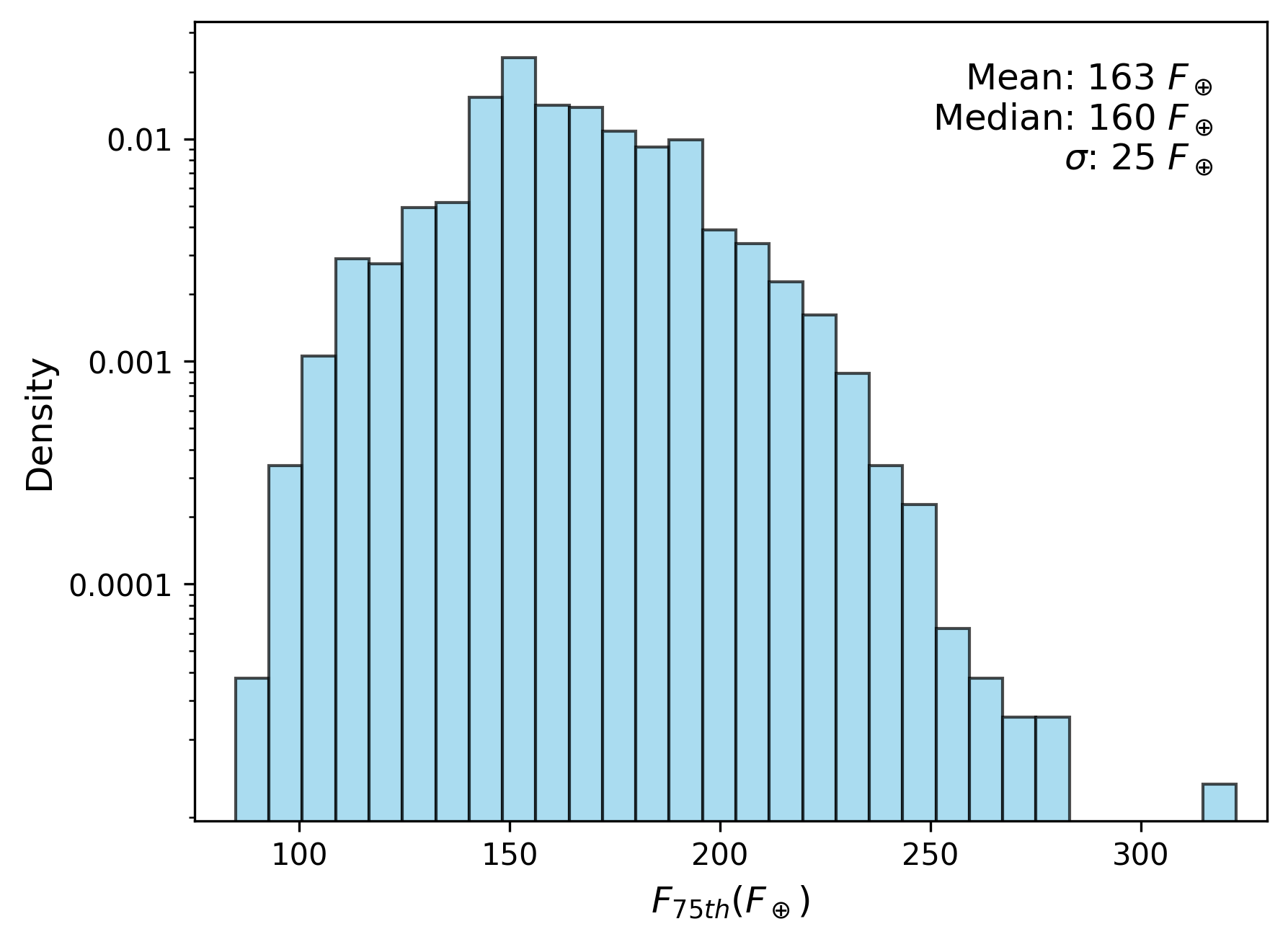}
    \caption{Histogram of the variable $F_{75\text{th}}$ for planets with $3\,R_\oplus \leq R_p \leq 11\,R_\oplus$, generated from 10,000 Monte Carlo simulations. Each simulation reduced the $S_1$ sample to the size of the $S_2$ and calculated the $75$th percentile of exoplanets with $3\,R_\oplus \leq R_p \leq 11\,R_\oplus$. The histogram is normalised to show the probability density.} 
    \label{fig:hist_F75}
\end{figure}

We performed $10,000$ Monte Carlo simulations, in each of which the $S_1$ sample (comprising $1527$ planets) was reduced to the size of the $S_2$ sample ($609$ planets) by randomly removing $918$ objects. In each simulation, we calculated the $F_{75\text{th}}$ of exoplanets with $3\,R_\oplus \leq R_p \leq 11\,R_\oplus$, whose distribution is shown in Fig. \ref{fig:hist_F75}.
We obtained a non-Gaussian $F_{75\text{th}}$ distribution with a mean value of $\approx 163\,F_\oplus$ and a standard deviation of $\approx 25\,F_\oplus$. The distribution shows a slight right-handed tail that is consistent with a skewness of $\approx 0.4$. The non-Gaussianity was also investigated using the Anderson-Darling (AD) test \citep{Anderson_Darling_1952}, whose statistics of $65$ exceeds the highest critical value by far ($1.092$) at the $1\%$ significance level. The Gaussianity null hypothesis is therefore rejected at all of the provided significance levels. 
We opted for the AD test instead of the more commonly used Shapiro-Wilk test since the latter is known to perform better for small sample sizes \citep{Shapiro1965}. 
We found that $99.84\%$ of our simulations yielded $F_{75\text{th}}$ values lower than $250\,F_\oplus$. Moreover, only one simulations out of $10,000$ returned a value of $F_{75\text{th}}\gtrsim 300\,F_\oplus$, which is still much lower than the $550\,F_\oplus$ we obtained for the $F$ -- $M_p$ diagram.

\subsubsection{Augmenting the $S_2$ sample}\label{sec:increase_S2}

In order to understand the effect of the small sample size on the definition of $F_{75\text{th}}$ and ultimately on the borders of the desert, we considered an augmented version of $S_2$ by assigning an approximate mass estimate, derived using \texttt{forecaster} \citep{Chen_2017}, to all objects in the $S_1$ sample without a measured planetary masses. We also included in the new enlarged sample planets with $\Delta M_p/M_p>30\,\%$. 
This procedure resulted in a $S_2$ sample of $1527$ confirmed planets, each with an estimated mass, regardless of the magnitude of the associated uncertainty. Nevertheless, even in this scenario, the $F_{75\text{th}}$ for planets with $10\,M_\oplus \leq M_p \leq 200\, M_\oplus$ remains slightly above $400\,F_\oplus$.

Thus, our statistical analysis suggests that the threshold value of the incident flux $F_{75\text{th}}$ is not affected by the poorer statistics in the flux-mass plane, but might be the result of a non-trivial physical effect. A similar result was obtained by \cite{Ma2021}: By employing a Bayesian mixture model, they found that the incident flux has a non-negligible effect on the $M_p$ -- $R_p$ relation. Future multivariate approaches that simultaneously take into account $(F, R_p, M_p)$ might shed light upon this issue.

\subsection{The mass-radius relation}
We also checked whether the GMM clustering processes performed in the two different spaces of parameters yielded any inconsistent results. In particular, one might wonder whether a planet of group $A_R$ (small radius) in the $F$-- $R_p$ diagram still belongs to group $A_M$ (low mass) in the $F$ -- $M_p$ space or if it falls within the domain of group $B_M$ (high mass). 
If this were not the case, it would mean that planets exist whose classification depends on the physical parameter used to describe them. 
We performed this analysis using the $S_3$ set of $512$ exoplanets with a well-constrained planetary mass and planetary radius measurements. Our results are listed below.
\begin{enumerate}
    \item We found that 271 planets ($\approx 53\%$ of $S_3$ sample) are classified as belonging to group $A$ according to both radius and mass (i.e. small radius $A_R$ and low mass $A_M$). We term this class $A_R A_M$ hereafter.
    \item We found that 29 planets are classified as belonging to group $B$ with respect to the radius and to group $A$ with respect to the mass (i.e. large radius $B_R$ and low mass $A_M$). We term this class $B_R A_M$ hereafter.
    \item  We found that one planet (K2-60 b; \citealt{Eigmuller2017}) was classified as belonging to group $A$ with respect to the radius and to group $B$ with respect to the mass (i.e. small radius $A_R$ and high mass $B_M$). We term this class $A_R B_M$ hereafter.
    \item We found that 211 planets ($\approx 41\%$ of $S_3$ sample) were classified as belonging to group $B$  for the radius and mass (i.e. large radius $B_R$ and high mass $B_M$). We term this class $B_R B_M$ hereafter.
\end{enumerate}

\begin{figure}
    \centering
\includegraphics[width=0.48\textwidth]{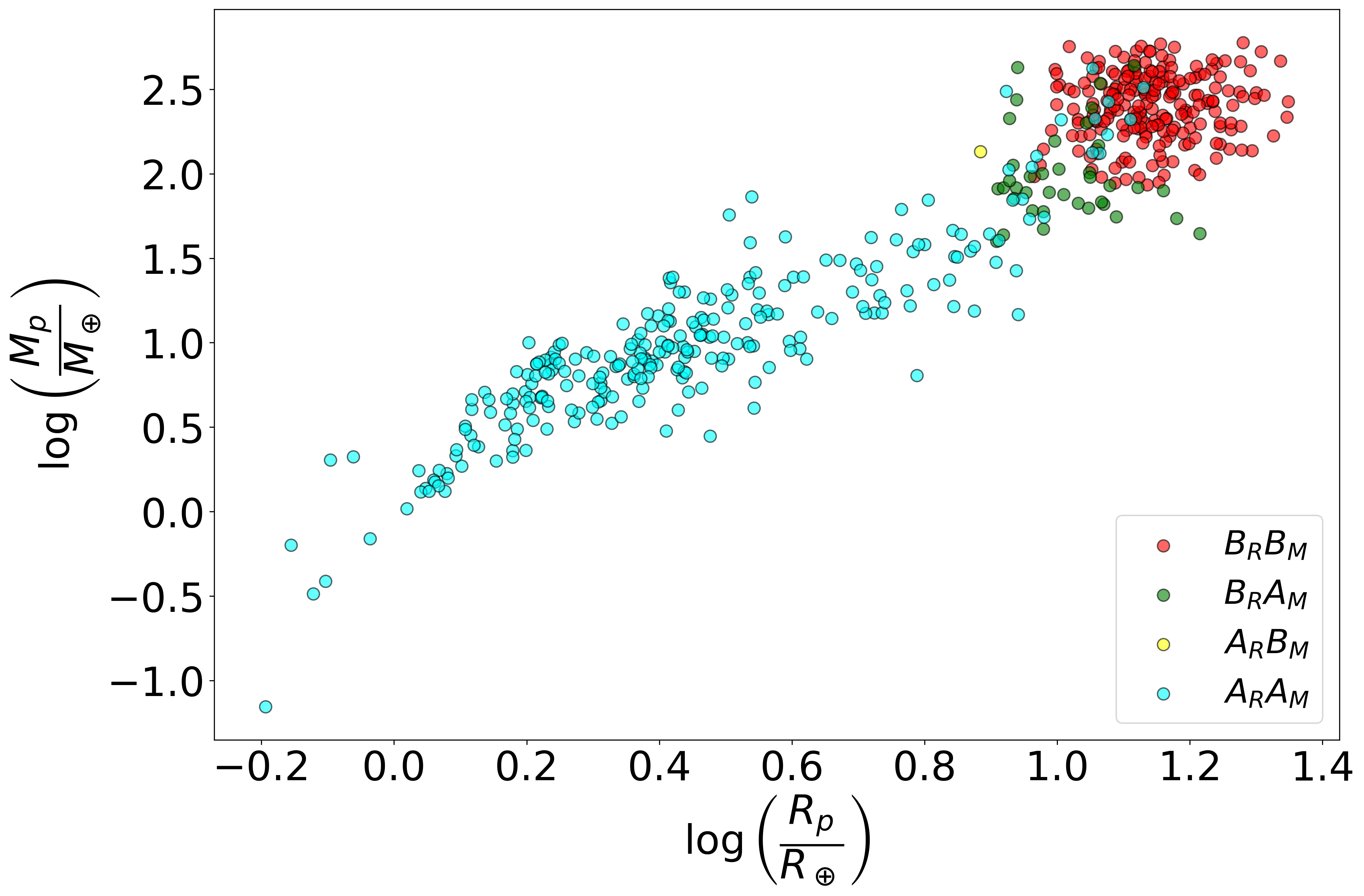}
    \caption{Distribution of $512$ objects of sample $S_3$ within the $\log R_p$ -- $\log M_p$ plane, coloured according to their classification with respect to the radius and mass. $271$ blue dots ($A_R A_M$) represent the planets classified as not giants for the radius and mass, $211$ red dots ($B_R B_M$) represent the planets labelled as giants for the radius and mass, $29$ green points ($B_R A_M$) are giant planets according to their radius and not giants according to their mass, and the one yellow point ($A_R B_M$) corresponds to a not-giant in terms of the radius and to a giant with respect to the planetary mass.}
    \label{fig:matrix_mass}
\end{figure}

In Fig. \ref{fig:matrix_mass} we show the distribution of $512$ objects within the $\log R_p$ -- $\log M_p$ plane, each coloured according to its classification with respect to both radius and mass. It is interesting to note that the $29$ planets classified as $B_R A_M$ belong to the transition of the $M (R)$ curve. 
The average mean planetary density of the $B_R A_M$ group is $\approx 0.68\,\text{g/}\text{cm}^3$ (approximately half of the mean Jupiter density) with a standard deviation of $\approx 0.61\,\text{g/}\text{cm}^3$. Thus, these objects might represent the well-known class of inflated hot Jupiters \citep{Sestovic2018}. 
We remark that some of the hybrid objects (i.e. belonging to groups $A_R B_M$ and $B_R A_M$) might simply be the result of the precision limit of the GMM clustering. 
K2-60 b \citep{Eigmuller2017} has a mean planetary density of $\approx 1.6\,\text{g/}\text{cm}^3$. Future spectroscopic follow-up investigations of these objects could shed light upon their proper position within the $\log R_p$ -- $\log M_p$ plane.

\subsection{Bayesian inference and model comparison with \textsc{Diamonds}}
\label{sec:diamonds_radius_inference}
In order to fit the boundaries in the $F$ -- $R_p$ plane that delimit the region of the desert, we exploited a series of different analytical models whose statistical significance was subsequently tested by means of a Bayesian model comparison process. 
All the fits were performed using the public Bayesian inference tool \textsc{Diamonds}\footnote{https://github.com/EnricoCorsaro/DIAMONDS} \citep{Corsaro2014}. \textsc{Diamonds} allowed us to simultaneously perform a Bayesian parameter estimation and model comparison using the nested-sampling Monte Carlo algorithm \citep{Skilling2004}, without limitations regarding the size of the dataset or the number of free parameters involved in the model. \textsc{Diamonds} returned the Bayesian evidence ($\mathcal{E}$) for each tested model. The Bayesian evidence represents the probability of the data in the light of a given model, that is, it averages the likelihood distribution over the parameter space set by the priors, namely by our choice of values of the free model parameters. In particular, $\mathcal{E}$ is defined as
\begin{equation}
\mathcal{E} \equiv \int_{\Sigma_\mathcal{M}} \mathcal{L} \left( \boldsymbol{\theta} \right)  \pi \left( \boldsymbol{\theta} \mid \mathcal{M} \right) d\boldsymbol{\theta} \, , 
\end{equation}
where $\boldsymbol{\theta}$ is the parameter vector, $\mathcal{L}$ is the likelihood, and $\pi$ is the prior on the free parameters as conditional to the choice of the model $\mathcal{M}$, the latter defining the parameter space $\Sigma_\mathcal{M}$ over which the integral is performed.
When we compare the two models $\mathcal{M}_1$ and $\mathcal{M}_2$, the corresponding Bayes factor $\mathcal{B}_{1,2}$ of $\mathcal{M}_1$ relative to  $\mathcal{M}_2$ is defined as the ratio of the model evidence, or equivalently, as $\ln \mathcal{B}_{1,2} \equiv \ln \mathcal{E}_1 - \ln \mathcal{E}_2$ when we adopt a logarithmic form. For $\ln \mathcal{B}_{1,2} \geq 5$ \citep{Jeffreys1961}, the evidence condition in favour of model $\mathcal{M}_1$ is strong, meaning that this model can be considered with a probability higher than 99\% to be the favoured choice.

The first model we took into account was a power-law model of the form
\begin{equation}
\frac{R_p}{R_{\oplus}} = \beta \left(\frac{F}{F_{\oplus}}\right)^{\alpha} \, ,
\end{equation}
with the free parameters $\alpha$ and $\beta$. 
In order to incorporate the uncertainties on the independent variable $F$, we linearised this model by adopting a logarithmic scale,
\begin{equation}
\ln \left( \frac{R_p}{R_{\oplus}} \right) = \alpha \ln \left(\frac{F}{F_{\oplus}}\right) + \ln \beta \, .
\end{equation}
This model, hereafter model $\mathcal{M}^R_1$, has the free parameters ($\alpha$, $\ln \beta$).

As a second option, we considered the linear model
\begin{equation}
\frac{R_p}{R_{\oplus}} = a \left(\frac{F}{F_{\oplus}}\right) + b \, ,
\end{equation}
hereafter termed model $\mathcal{M}^R_2$, with ($a$, $b$) as free parameters.

We performed a Bayesian inference on the two models presented above by adopting a Gaussian likelihood that incorporated the uncertainties on the planet radii and fluxes. The corresponding log-likelihood reads
\begin{equation}
\Lambda (\boldsymbol{\theta}) = \, \Lambda_0 (\boldsymbol{\theta}) \, - \frac{1}{2} \sum^N_{i=1} \, \left[ \frac{\Delta_i (\boldsymbol{\theta})}{\widetilde{\sigma_i} (\boldsymbol{\theta})} \right]^2 \, ,
\label{eq:likelihood}
\end{equation}
where $\boldsymbol{\theta}$ is the parameters vector (e.g. $(\alpha, \ln \beta)$ for model $\mathcal{M}_{1}$), $N$ is the total number of points considered in each boundary, $\widetilde{\sigma_i} (\boldsymbol{\theta})$ is the relative uncertainty over $\boldsymbol{\theta}$, and $\Lambda_0 (\boldsymbol{\theta})$ is a term depending on the relative uncertainties, given by
\begin{equation}
\Lambda_0 (\boldsymbol{\theta}) = - \sum^N_{i=1} \ln \sqrt{2 \pi} \widetilde{\sigma_i} (\boldsymbol{\theta}) \, .
\end{equation}

Lastly, we considered a third model, termed $\mathcal{M}^R_3$, which extended the linearity to a quadratic regime, but this time, without the uncertainties in the covariates $F$. 
The model reads
\begin{equation}
\frac{R_p}{R_{\oplus}} = \tilde{a} \left(\frac{F}{F_{\oplus}}\right) + \tilde{b} \left(\frac{F}{F_{\oplus}}\right)^2 + \tilde{c} \, ,
\end{equation}
and the free parameters are ($\tilde{a}$, $\tilde{b}$, $\tilde{c}$). The outputs of  \textsc{Diamonds} in the flux-radius diagram are discussed in Sect. \ref{sec:diamonds_rad}.

In addition to the investigation of the $F$ -- $R_p$ plane, we performed a similar analysis in the $F$ -- $M_p$ space. In particular, we analysed the analogous of models $\mathcal{M}^R_1$ and $\mathcal{M}^R_2$ by replacing the planet radius $R_p$ with the planet mass $M_p$, from which we obtained models $\mathcal{M}^M_1$ and $\mathcal{M}^M_2$. We did not consider the quadratic model $\mathcal{M}^R_3$ because too few points are available to constrain the larger number of free parameters.
The outputs of  \textsc{Diamonds} in the flux-mass diagram are discussed in Sect. \ref{sec:diamonds_mass}.

\section{Results}
\label{sec:results}

The analysis presented in Sect. \ref{sec:methods} shows that a paucity of highly irradiated Neptunes and sub-Jovians exists in the flux-radius and flux-mass diagrams. This feature suggests a dichotomy of planets at a high level of incident flux ($\gtrsim 200\,F_\oplus$): sub-Neptunes/super-Earths, and Jupiters/super-Jupiters. 
By using the GMM algorithm, we determined the 
 upper and lower edges in the flux-radius and flux-mass planes as described in Sects. \ref{sec:flux_rad_analysis} and \ref{sec:flux_mass_analysis}. 
We then fitted each edge in the two planes using \textsc{Diamonds}, as shown in Sects. \ref{sec:diamonds_radius_inference} and \ref{sec:diamonds_radius_inference}. 
Thus, the depletion region delimited by the two boundaries here takes the orbital period of the planet and the fundamental properties of the host star into account.

\subsection{Flux-radius plane}
\label{sec:diamonds_rad}

The Bayesian evidence $\mathcal{E}$ computed through \textsc{Diamonds} is listed in Table~\ref{tab:ev}. The model that is clearly favoured for the lower and upper boundary is the power-law model $\mathcal{M}^R_1$, for which the Bayes factors largely exceeds the strong-evidence condition ($\ln \mathcal{B} > 5.0$).
We noted that the model involving a quadratic dependence upon $F$ is subject to significant departures from the data regime for values outside the range imposed by the data. 
As appears evident from the reported Bayesian evidence, the larger number of free parameters of this latter model significantly penalises its statistical weight as compared to the other models that we took into account in our study \citep{Jeffreys1961}. 

\begin{table}
\small
\centering
 \caption{Bayesian evidence for the models defined in this work.}
\begin{tabular}{lrr}
  \hline
  \\[-8pt]
 Model & \multicolumn{1}{c}{$\ln \mathcal{E}$ (Upper Boundary)} & \multicolumn{1}{c}{$\ln \mathcal{E}$ (Lower Boundary)}\\[1pt]
  \hline
  \\[-8pt]
  \rowcolor[gray]{0.9} $\mathcal{M}^R_1$ & 6.4655 & -6.7525 \\[1pt]
  $\mathcal{M}^R_2$ & -16.6151 & -44.4909 \\[1pt]
  $\mathcal{M}^R_3$ & -25.7549 & -69.6771 \\[1pt]
  \rowcolor[gray]{0.9} $\mathcal{M}^M_1$ & -3.4143 & -4.4521 \\[1pt]
  $\mathcal{M}^M_2$ &  -26.0992 & -15.7327 \\[1pt]
  \hline
 \end{tabular}
\label{tab:ev}
\tablefoot{The models favoured in the model comparison process is highlighted with a grey band.}
\end{table}

In Table \ref{tab:m1} we report the best-fitting values for $\alpha$ and $\ln\beta$ of the power-law model $\mathcal{M}^R_1$ as retrieved by \textsc{Diamonds}. We name the area comprised between the boundaries obtained with the $\mathcal{M}^R_1$ model irradiation desert in the $F$ -- $R_p$ plane.

\begin{table}
\small
\centering
\caption{Parameter estimation results for the power-law models $\mathcal{M}^R_1$ and $\mathcal{M}^M_1$ as obtained by \textsc{Diamonds}.}
\begin{tabular}{lrr}
  \hline
  \\[-8pt]
 Model & \multicolumn{1}{c}{$\alpha$} & \multicolumn{1}{c}{$\ln \beta$}\\[1pt]
  \hline
  \\[-8pt]
 $\mathcal{M}^R_1$ (Lower boundary) & $-0.27_{-0.02}^{+0.02}$ & $2.64_{-0.10}^{+0.11}$ \\[4.5pt]
 $\mathcal{M}^R_1$ (Upper boundary) & $0.11_{-0.03}^{+0.02}$ & $1.56_{-0.15}^{+0.17}$ \\[4.5pt] 
  $\mathcal{M}^M_1$ (Lower boundary) & $-0.70_{-0.13}^{+0.16}$ & $7.82_{-1.16}^{+1.03}$ \\[4.5pt]
 $\mathcal{M}^M_1$ (Upper boundary) & $0.47_{-0.22}^{+0.19}$ & $1.78_{-1.49}^{+1.77}$ \\[4.5pt] 
  \hline
 \end{tabular}
\label{tab:m1}
\tablefoot{Median values and corresponding 68.3\,\% Bayesian credible limits are indicated.}
\end{table}

\begin{figure*}
    \centering
    \includegraphics[width=0.9\textwidth]{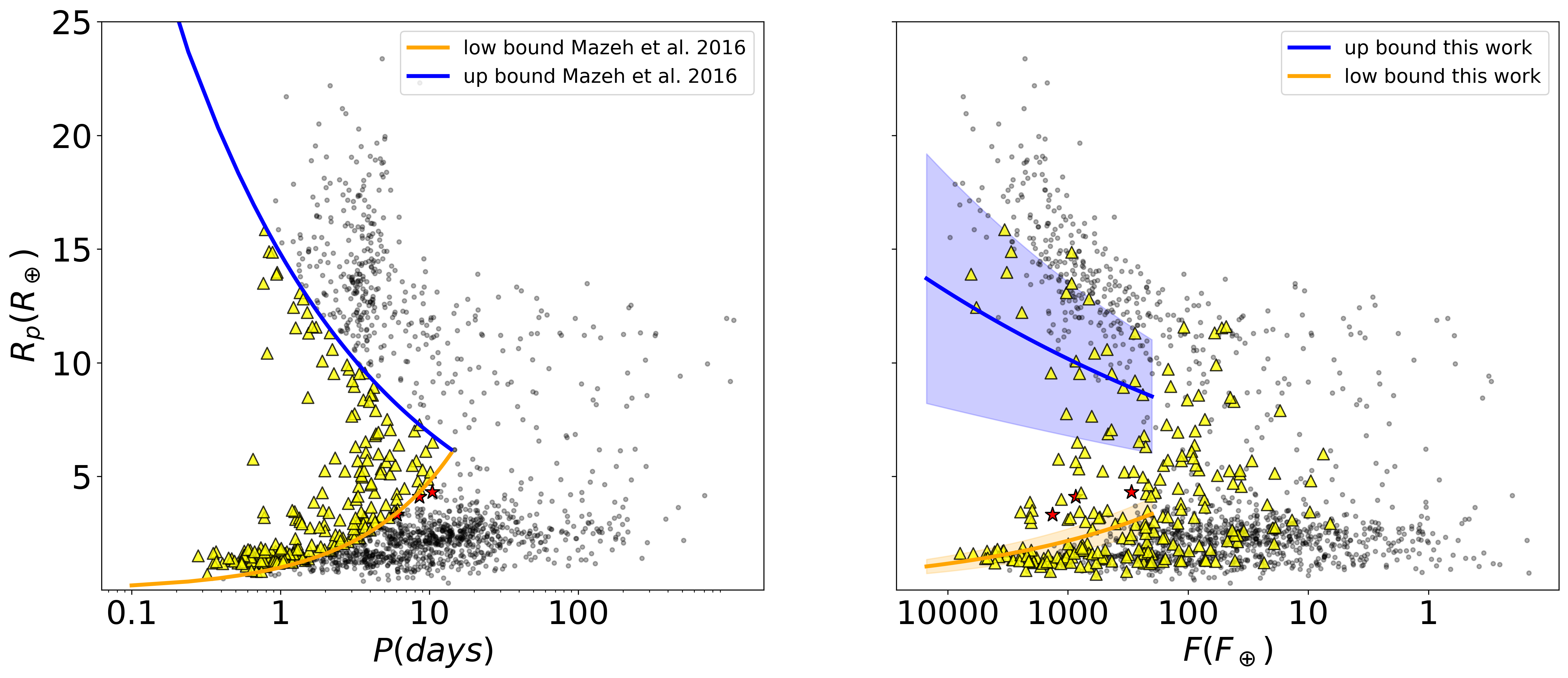}
    \caption{Comparison of the $S_1$ sample distribution in the $P$ -- $R_p$ and $F$ -- $R_p$ planes to highlight the irradiation desert. \textit{Left}: Distribution of the $S_1$ sample in the $P$ -- $R_p$ diagram along the edges obtained by \citealt{Mazeh2016}. The yellow triangles represent the 221 hot Neptunes within the $S_1$ sample according to the definition of \citealt{Mazeh2016} in the $P$ -- $R_p$ plane. The black dots represent all the planets that lie outside the desert. \textit{Right}: Distribution of the $S_1$ sample in the $F$ -- $R_p$ plane. The orange and blue lines represent the lower and upper boundary obtained in Sect. \ref{sec:diamonds_rad} alongside their confidence bands ($1\,\sigma$). The yellow triangles show the $194$ out of $221$ objects that do not lie within the irradiation desert defined in this work. The yellow triangles used to designate the hot Neptunes in the $P$ -- $R_p$ diagram still refer to the same planets in the right plot. The red stars indicate planets that fall within our defined boundaries but not within those outlined by \citealt{Mazeh2016}.}
    \label{fig:Magliano_Mazeh_radius}
\end{figure*}

In Fig. \ref{fig:Magliano_Mazeh_radius} we show the distribution of $S_1$ sample in the $P$ -- $R_p$ diagram alongside the \cite{Mazeh2016} boundaries and within the $F$ -- $R_p$ diagram alongside the lower and upper boundaries obtained in this work. Out of $221$ hot Neptunes, identified in the $P$ -- $R_p$ diagram, $194$ fall outside the irradiation desert in the $F$ -- $R_p$ plane or lie along its edges. 
Future refinements of the orbital and physical parameters of the objects lying on the edges could shed light upon their membership of the desert. Only $27$ out of $221$ planets occupy the statistical scarcely populated region within the $F$ -- $R_p$ diagram. Thus, according to this new perspective, the $194$ planets are not so rare, even though they were earlier classified as hot Neptunes. 
Furthermore, we also checked whether any planet of our sample that resides outside the desert as defined by \cite{Mazeh2016} (in the $P$ -- $R_p$ diagram) would fall inside the desert derived in this work. Only three planets satisfy this condition, and they are Kepler-1270 b ($R_p=3.32^{+0.30}_{-0.65}\,R_\oplus$ and $P=6.03356196 \pm 0.00006551$ days; \citealt{Morton2016}), Kepler-815 b ($R_p=4.11^{+0.19}_{-0.33}\,R_\oplus$ and $P=8.57503552\pm 0.0000591$ days)  and KOI-94 c ($R_p=4.32\pm 0.41\,R_\oplus$ and $P=10.423707\pm 0.000026$ days; \citealt{Weiss2013,Masuda2013}). As shown in  Fig. \ref{fig:Magliano_Mazeh_radius}, while these planets lie nearby but outside of the lower boundary as defined by \cite{Mazeh2016}, they fall well inside the desert found in this work.

\subsection{Flux-Mass plane}
\label{sec:diamonds_mass}

We also fitted the two edges of the sub-Jovian and Neptune desert within the $F$ -- $M_p$ plane using \textsc{Diamonds}.
As in the previous case, the model that is clearly favoured for the lower and upper boundary is the power-law model $\mathcal{M}^M_1$, and its Bayes factors largely exceeds the strong-evidence condition ($\ln \mathcal{B} > 5.0$), as shown in Table \ref{tab:ev}.

\begin{figure*}
    \centering
    \includegraphics[width=0.9\textwidth]{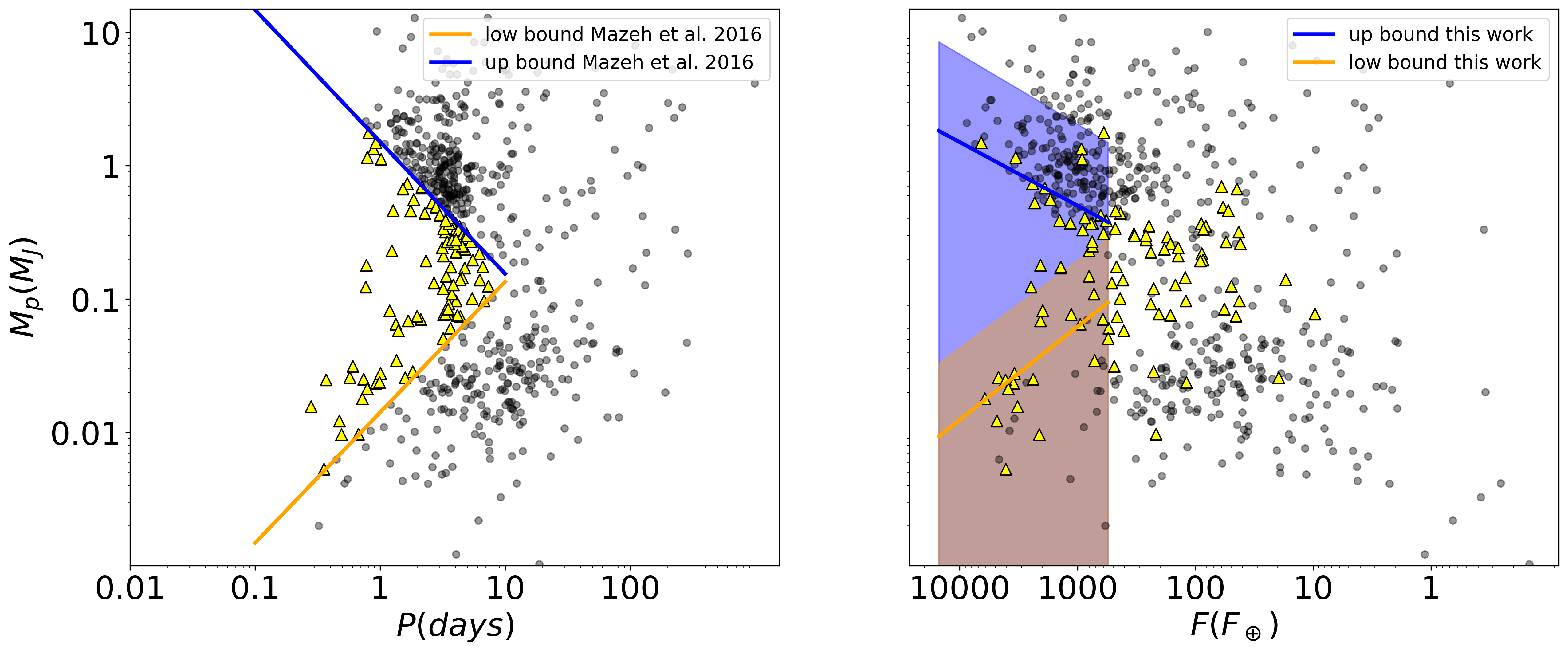}
    \caption{Comparison of the $S_3$ sample distribution in the $P$ -- $M_p$ and $F$ -- $M_p$ planes to highlight the irradiation desert. \textit{Left}: Distribution of the $S_3$ sample in the $P$ -- $M_p$ diagram along the edges obtained by \citealt{Mazeh2016}. The yellow dots represent the 110 hot Neptunes within $S_3$ according to the definition of \citealt{Mazeh2016} in the $P$ -- $M_p$ plane. The black dots represent planets outside the desert. \textit{Right}: Distribution of our sample in the $F$ -- $M_p$ plane. The orange and blue lines represent the lower and upper boundary obtained in Sect. \ref{sec:diamonds_mass} alongside their confidence bands ($1\,\sigma$). The red dots show that $63$ out of $110$ objects experience an incident flux lower than $550\,F_\oplus$. The yellow triangles used to designate the hot Neptunes in the $P$ -- $M_p$ diagram still refer to the same planets in the right plot.}
    \label{fig:Magliano_Mazeh_mass}
\end{figure*}

In Table \ref{tab:m1} we report the best-fitting values for $\alpha$ and $\ln\beta$ of the power-law model $\mathcal{M}^M_1$ as retrieved by \textsc{Diamonds}. We name the area between the boundaries obtained with the $\mathcal{M}^M_1$ model the irradiation desert in the $F$ -- $M_p$ plane.
With respect to the results obtained for the flux-radius plane (see Sect. \ref{sec:diamonds_rad}), the  errors on the fitted parameters are relatively larger because the statistics in the flux-mass diagram are larger, as already discussed in Sect.  \ref{sec:flux_mass_analysis}.
Within the $S_3$ sample, $110$ planets are found to populate the desert obtained by \citealt{Mazeh2016} in the $P$ -- $M_p$ diagram. 
Figure \ref{fig:Magliano_Mazeh_mass} depicts the distribution of these planets in the $F$ -- $M_p$ diagram, together with the lower and upper boundaries obtained in this work. 
As discussed in Sect.  \ref{sec:flux_mass_analysis}, because the statistics in the flux-mass plane are lower, the confidence bands of the edges overlap when we take the uncertainties on the fit in the flux-mass plane into account, as shown in Fig. \ref{fig:Magliano_Mazeh_mass}. 
However, 63 planets out of 110 still receive an incident flux $F\leq 550\,F_\oplus$ and therefore do not populate the irradiation desert in the $F$ -- $M_p$ plane. This might be the result of using a different $F_{75\text{th}}$ between the $F$ -- $R_p$ and $F$ -- $M_p$ planes because this effect does not clearly arise when we considered the orbital period in place of the incident flux.

In Sect. \ref{sec:increase_S2} we constructed an augmented version of  sample $S_2$ by using \texttt{forecaster} to obtain an $M_p$ estimate for planets without mass measurements
and keeping objects regardless of the magnitude of the uncertainty $\Delta M_p/M_p$. 
In principle, we could have used this sample to compute the analytical expression of the lower and upper boundary in the $F$ -- $M_p$ plane. However, even though this approach would yield a larger data set to be fitted with \textsc{Diamonds}, we definitely lose much in accuracy. Typically, the uncertainties in $M_p$ estimated by \texttt{forecaster} range from approximately $30\,\%$ to $50\,\%$ for well-characterised exoplanets, but this can be significantly higher for planets with less precise data. 
In particular, for our data set, we found the uncertainties to be normally distributed around $\approx 60\,\%$ with a standard deviation of about $5\,\%$. Consequently, even though employing an augmented sample might help us to mitigate the issues associated with limited statistics in the $F$ – $M_p$ plane, the results would ultimately rely on a highly uncertain sample, which would lead to unreliable boundary points.

\subsection{Planetary mean density}

In Sect. \ref{sec:diamonds_rad} and \ref{sec:diamonds_mass} we obtained the analytic expressions of the boundaries of the sub-Jovian and Neptune desert in the $F$ -- $R_p$ and $F$ -- $M_p$ planes. For each edge, by combining  $R_p(F)$ and $M_p(F)$, we can obtain a mass-radius relation in order to understand what types of planet populate them. 
In particular, the mass-radius relation for the upper boundary is given by
\begin{equation}
    \dfrac{R_p}{R_\oplus}=(3.1\pm2.5)\left(\dfrac{M_p}{M_\oplus}\right)^{0.23\pm 0.15} \, ,
    \label{eq:radius_mass_up}
\end{equation}
while the mass-radius relation describing the lower boundary is

\begin{equation}
    \dfrac{R_p}{R_\oplus}=(0.7\pm 0.7) \left(\dfrac{M_p}{M_\oplus}\right)^{0.39\pm 0.07} \, .
    \label{eq:radius_mass_low}
\end{equation}

\begin{figure*}
    \centering
    \includegraphics[width=0.9\textwidth]{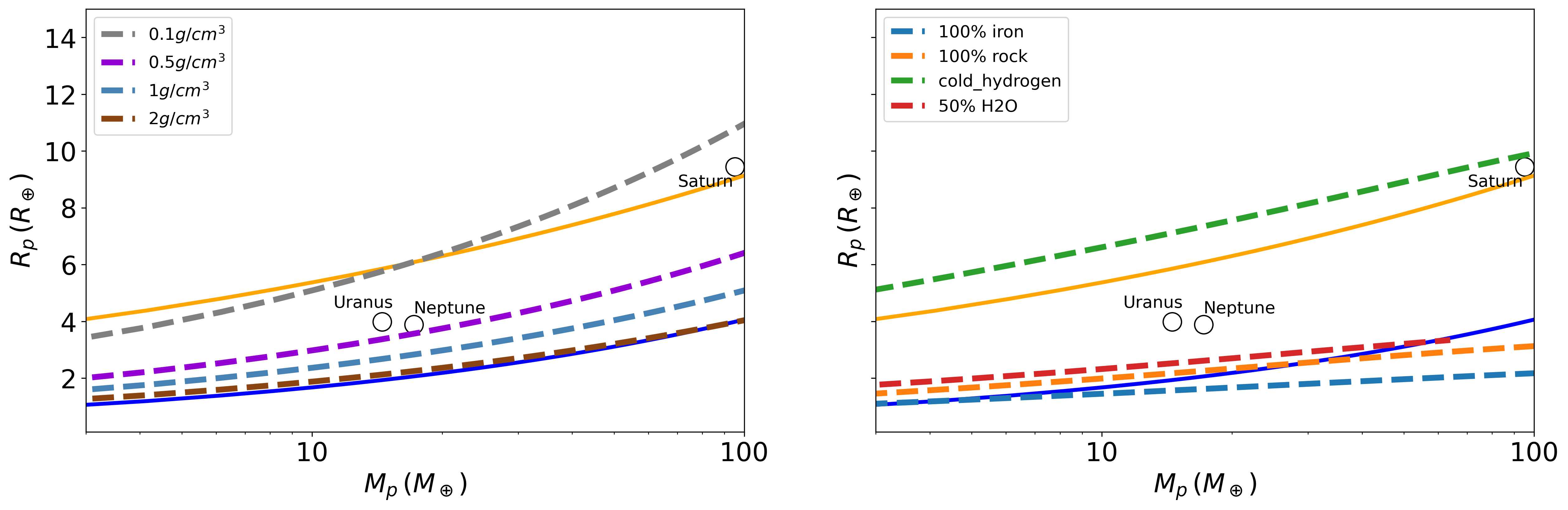}
    \caption{Mass-radius relations for the upper (orange line) and lower boundary (blue line) given by Eqs. \eqref{eq:radius_mass_up} and \eqref{eq:radius_mass_low}, respectively. We present different curves at fixed density (left). While the upper boundary is consistent with a very low mean planetary density ($\approx 0.1\,\text{g/}\text{cm}^3$), the lower boundary comprises different values spanning from $\approx 1\, \text{g/}\text{cm}^3$ to $\approx 2\, \text{g/}\text{cm}^3$. We also present different mass-radius curves for different planetary compositions (right).}
    \label{fig:mean_density}
\end{figure*}

We show the mass-radius relations for the two boundaries in Fig. \ref{fig:mean_density}. 
Planets populating the upper edge of the sub-Jovian and Neptune desert in the $F$ -- $R_p$ and $F$ -- $M_p$ planes are compatible with very low planetary density values ($\approx 0.1 \,\text{g/}\text{cm}^3$), thus suggesting a gaseous nature. The resulting exponent $(0.23 \pm 0.15)$ agrees well with the findings of \cite{Mazeh2016} within the numerical uncertainties.
On the other hand, planets lying on the lower edge systematically have a higher density. The spread of the curve, which is mostly due to the uncertainty on the mass estimates, leads to a certain level of ambiguity: These planets could either be covered by a gaseous envelope that drastically reduces the mean planetary density, or they might be silicate worlds. The former case might be compatible with a low-mass planet covered by a H/He envelope that undergoes photoevaporation due to the stellar irradiation of the host. The latter might represent the final stage of the previous scenario where the planet has lost part of its envelope.

\section{Conclusions}
\label{sec:conclusions}

We have expanded upon the seminal work of \cite{Mazeh2016} by studying the sub-Jovian and Neptune desert in more detail. This is 
the region of the $P$ -- $R_p$ and $P$ -- $M_p$ space parameters from which Neptunian exoplanets with short orbital periods are notably absent. The orbital period is not a complete physical variable for understanding the nature of this phenomenon because it does not account for stellar radiation. Our analysis therefore incorporated the incident flux from the host star in order to provide a more comprehensive picture of the factors influencing the formation and evolution of these intriguing exoplanets.

Our  analysis revealed that the incident flux plays an important role in shaping the boundaries of the sub-Jovian and Neptune desert. We analysed the dependences of the upper and lower edges on the incident flux in the flux-radius and flux-mass space parameters. We employed \textsc{Diamonds}, a novel Bayesian nested-sampling tool, to infer the best models that fit the two boundaries in both planes. 
Even when we accounted for the incident flux, we still found a paucity of known Neptune- and Saturn-like exoplanets experiencing high-intensity levels of stellar radiation. We called the region in which they lie the irradiation desert. Moreover, $\approx 87\,\%$ of planets in the desert defined by \cite{Mazeh2016} in the period-radius plane do not lie within the irradiation desert we defined in the flux-radius diagram. A similar result was obtained in the flux-mass plane. Thus, the sub-Jovian and Neptune desert is likely much more desolate than we would expect. We stress that our method yielded boundaries that delimit a region that admits contamination by a few moderately irradiated Neptunes and sub-Jovians.

We also retrieved the mass-radius relation for the planets lying close to the two boundaries. 
While the upper boundary is populated by exoplanets with a very low mean planetary density, which is compatible with gas giants that are stable against the photoevaporation scenario, the lower boundary comprises a broader range of planetary types: from planets with gaseous envelops to objects with a higher mean planetary density, which might be compatible with worlds dominated by silicates.

Future observations with advanced telescopes such as the James Webb Space Telescope (JWST, \citealt{Gardner2023}) and exoplanetary missions, as well as an ongoing analysis of the thousands of planet candidates that have yet to be confirmed, will help us to refine the calculations of the two boundaries and will thus allow us to potentially reveal new insights into the nature and origins of Neptune-sized exoplanets in close orbits. 
In particular, the PLAnetary Transits and Oscillations of stars (PLATO, \citealt{Rauer2014,Rauer2024}) mission is expected to discover several hundreds to thousands of Neptune-sized exoplanets orbiting their star in fewer than $5$ days, depending on the assumed planet occurrence rates \citep{Matuszewski2023}. PLATO will not only enhance our understanding of planetary architectures by expanding the exoplanetary populations, but its high-photometric precision, combined with a host star characterisation via asteroseismology, will allow us to refine the planetary and stellar parameters. This will reduce the uncertainty on the data used in this work.

As the volume of these discoveries increases, there will be a critical need for efficient vetting processes, both through human expertise and automated machine-learning techniques (e.g. \citealt{McCauliff2015,Coughlin2016,Mislis2016,Shallue2018,Yu2019,Tey2023,Fiscale2023}), to ensure the accuracy and reliability of confirmed planets. However, the complexity of machine-learning models, which are often perceived as black-box systems, can limit transparency and scientific understanding. Therefore, integrating explainable machine-learning methods \citep{Maratea2019} into future vetting pipelines will be essential. These techniques will allow scientists to maintain control over and insight into the decision-making process, enabling greater confidence in the discoveries and fostering collaboration between computational tools and human expertise in the validation of new exoplanet candidates.
\begin{acknowledgements}

      This research has made use of the NASA Exoplanet Archive, which is operated by the California Institute of Technology, under contract with the National Aeronautics and Space Administration under the Exoplanet Exploration Program.
      \newline
      We thank the anonymous referee whose careful reading helped to improve the quality of our manuscript.
      \newline
      E.C. acknowledges support from PLATO ASI INAF agreement no. 2022-28-HH.0 “PLATO Fase D”.
      \newline
      G.C. thanks the University of Napoli Federico II (project: FRA-CosmoHab, CUP E65F22000050001).
      \newline
\textit{Software}: \textsc{Diamonds} \citep{Corsaro2014}, \texttt{forecaster} \citep{Chen_2017}.
\end{acknowledgements}

\bibliographystyle{aa}
\bibliography{biblio.bib}

\end{document}